\documentclass[12pt,reqno]{amsart}

\usepackage{latexsym}

\newcommand{\szego}{Szeg\"o }
\newcommand{\nhat}{\raisebox{2pt}{$\wh{\ }$}}

\newcommand{\inv}{^{-1}}
\newcommand{\kahler}{K\"ahler }
\newcommand{\sqrtn}{\sqrt{N}}
\newcommand{\wt}{\widetilde}
\newcommand{\wh}{\widehat}
\newcommand{\PP}{{\mathbb P}}
\newcommand{\R}{{\mathbb R}}
\newcommand{\C}{{\mathbb C}}

\newcommand{\Z}{{\mathbb Z}}
\newcommand{\B}{{\mathbb B}}
\newcommand{\CP}{\C\PP}
\renewcommand{\d}{\partial}
\newcommand{\dbar}{\bar\partial}
\newcommand{\ddbar}{\partial\dbar}
\newcommand{\D}{{\mathbf D}}
\newcommand{\E}{{\mathbf E}\,}
\renewcommand{\H}{{\mathbf H}}
\newcommand{\half}{{\frac{1}{2}}}
\newcommand{\vol}{{\operatorname{Vol}}}

\newcommand{\FS}{{{\operatorname{FS}}}}

\renewcommand{\phi}{\varphi}

\newcommand{\ccal}{\mathcal{C}}
\newcommand{\dcal}{\mathcal{D}}
\newcommand{\ecal}{\mathcal{E}}
\newcommand{\fcal}{\mathcal{F}}
\newcommand{\gcal}{\mathcal{G}}
\newcommand{\hcal}{\mathcal{H}}
\newcommand{\lcal}{\mathcal{L}}

\newcommand{\ocal}{\mathcal{O}}
\newcommand{\scal}{\mathcal{S}}
\newcommand{\ucal}{\mathcal{U}}

\newcommand{\jcal}{\mathcal{J}}
\newcommand{\al}{\alpha}

\newcommand{\ga}{\gamma}

\newcommand{\La}{\Lambda}
\newcommand{\la}{\lambda}
\newcommand{\ep}{\varepsilon}
\newcommand{\de}{\delta}
\newcommand{\De}{\Delta}
\newcommand{\om}{\omega}

\newcommand{\bbb}{|\!|\!|}
\newtheorem{theo}{{\sc Theorem}}[section]
\newtheorem{cor}[theo]{{\sc Corollary}}
\newtheorem{lem}[theo]{{\sc Lemma}}

\newenvironment{rem}{\medskip\noindent{\it Remark:\/} }{\medskip}
\newenvironment{defin}{\medskip\noindent{\it Definition:\/} }{\medskip}

\title{Universality and Scaling of Zeros on Symplectic Manifolds}

\author{Pavel Bleher}
\address{Department of Mathematical Sciences, IUPUI, Indianapolis, IN
46202,
USA}
\email{bleher@math.iupui.edu}
\author{Bernard Shiffman}
\address{Department of Mathematics, Johns Hopkins University, Baltimore,
MD
21218, USA}
\email{shiffman@math.jhu.edu}
\author{Steve Zelditch}
\address{Department of Mathematics, Johns Hopkins University, Baltimore,
MD
21218, USA}
\email{zelditch@math.jhu.edu}

\thanks{Research partially supported by NSF grants
\#DMS-9970625 (first author), \#DMS-9800479 (second author),
\#DMS-9703775
(third author).}

\begin{document}

\begin{abstract} 

This article is concerned with random holomorphic polynomials and their
generalizations to algebraic and symplectic geometry.  A natural
algebro-geometric generalization involves random
holomorphic sections $H^0(M,L^N)$ of the powers of any positive line bundle $L
\to M$ over any complex manifold.  Our main interest is in the statistics of
zeros of $k$ independent sections (generalized polynomials) of degree $N$ as
$N\to\infty$.  We fix a point $P$ and focus on the ball of radius $1/\sqrt{N}$
about $P$.  Magnifying the ball by the factor $\sqrt{N}$, we found in a
prior work that the
statistics of the configurations of simultaneous zeros of random $k$-tuples of
sections tend to a universal limit independent of $P,M,L$. We review this
result and generalize it further to the case of pre-quantum line bundles over
almost-complex symplectic manifolds $(M,J,\omega)$.  Following \cite{SZ2}, we
replace $H^0(M,L^N)$ in the complex case with the ``asymptotically
holomorphic'' sections defined by Boutet de Monvel-Guillemin and (from another
point of view) by Donaldson and Auroux.  We then give a generalization to an
$m$-dimensional setting of the Kac-Rice formula for zero correlations, which
we use together with the results of \cite{SZ2} to prove that the scaling
limits of the $n$-point correlation functions for zeros of random $k$-tuples
of asymptotically holomorphic sections belong to the same universality class
as in the complex case.  In our prior work, we showed that the limit
correlations are short range; here we show further that the limit ``connected
correlations'' decay exponentially with respect to the square of the maximum
distance between points.
\end{abstract}

\maketitle

\section{Introduction}  A well-known theme in random matrix theory
(RMT), zeta functions,   quantum chaos, and statistical mechanics,   is
the universality of scaling limits of correlation functions.
In RMT, the relevant correlation functions are for eigenvalues
of  random matrices (see \cite{D,TW,BZ,BK,So} and
their references).
In the case of zeta functions, the correlations  are between the zeros
\cite{KS}.  In quantum
dynamics,  they are between  eigenvalues
of `typical'  quantum maps whose underlying classical maps have a specified
dynamics.  In the `chaotic case' it
is conjectured that the correlations should belong to the  universality
class of RMT, while in integrable cases they
should belong to that of Poisson processes. The latter has been confirmed
for certain families of integrable quantum maps, scattering matrices  and
Hamiltonians (see \cite{Ze.2,RS,Sa,ZZ} and their references).  In
statistical mechanics, there is a large
literature
on universality of critical exponents \cite{CAR}; other
rigorous results include analysis of universal scaling limits of Gibbs
measures at critical points \cite{SI}.  In this article we are concerned
with a somewhat new arena for scaling and universality,
namely that of RPT (random polynomial theory) and its algebro-geometric
generalizations \cite{Ha,Halp,BBL,BD,BSZ1,BSZ2,SZ,NV}.  The focus of these
articles is on the configurations and correlations of zeros of random
polynomials and their generalizations, which we discuss below. Random
polynomials can also be used to define random holomorphic maps to projective
space, but
we leave that for the future.  Our purpose here is partly to  review the
results of
\cite{SZ,  BSZ1,BSZ2} on universality of scaling limits of correlations 
between  zeros of random holomorphic sections on complex manifolds.
More significantly, we give an improved version of our formula from
\cite{BSZ2} for determining zero correlations from joint probability
distributions, and we apply this formula together with results in 
\cite{SZ2} to extend our limit zero correlation formulas to the case of
almost-complex symplectic manifolds.

Notions of universality depend on  context.  In RMT, one fixes a set of
matrices (e.g. a group $U(N)$ or a symmetric space such as $Sym(N)$, the 
$N\times N$
real symmetric matrices) and endows it with certain kinds of probability
measures $\mu_N$.  These measures
bias towards certain types of matrices and away from others, and one may ask
how the
eigenvalue correlations depend on the $\{\mu_N\}$ in the large $N$ limit.
In RPT one could similarly endow spaces of polynomials of degree $N$ with a
variety of measures, and ask how correlations between zeros depend on them
in the large $N$ limit.  However, the version of universality which concerns
us  in this article and in \cite{BSZ1,BSZ2} lies in another direction.
We are interested in very general notions of {\it polynomial} that arise in
geometry, and in how the
statistics of  zeros depends on the geometric setting in which these
polynomials live.  We will always endow our generalized polynomials with
Gaussian
measures (or with essentially equivalent spherical measures).

The generalized polynomials studied in \cite{SZ, BSZ1, BSZ2}  were
holomorphic sections $H^0(M, L^N)$  of powers of a positive line bundle $L
\to M$ over a compact K\"ahler manifold $(M, \omega)$ of a given dimension
$m$. Such sections form the Hilbert space of quantum wave functions which
quantize $(M, \omega)$ in the sense of geometric quantization \cite{At,W}.
Recall that geometric quantization begins with a symplectic manifold $(M, 
\omega)$
such that $\frac{1}{\pi}[\omega] \in H^2(M, \Z).$ There then exists a 
complex
hermitian line bundle $(L, h) \to M$ and a hermitian connection $\nabla$ 
with
curvature $\omega.$  To obtain a Hilbert space of sections, one needs 
additionally
to fix a {\it polarization} of $M$, i.e. a Lagrangean sub-bundle $\mathcal 
L$ of $TM$, and we
define {\it polarized sections\/} to be those satisfying $\nabla_v s = 0$ 
when
$v$ is tangent to $\mathcal L$.  In the K\"ahler
case, one takes  $\mathcal L = T^{1,0}M$, the holomorphic sub-bundle.  Thus,
polarized sections are holomorphic sections.  The power $N$ plays the role 
of the
inverse Planck constant, so that the high power $N \to \infty$ limit is the 
semiclassical
limit.

When
$M =\CP^m$ and $L = \mathcal O(1)$ (the hyperplane section line
bundle), holomorphic sections of $L^N = \mathcal O(N)$ are just
homogeneous holomorphic
polynomials of degree $N$.  (In general, one may embed $M \subset \CP^d$ for
some $d$, and then holomorphic sections $s \in H^0(M, L^N)$ may  be
identified with restrictions of polynomials
on $\CP^d$ to $M$, for $N\gg 0$ by the Kodaira vanishing theorem.) We equip
$L$ with a hermitian metric
$h$ and endow
$M$ with
the volume form $dV$ induced by the curvature $\omega$ of $L$.  The pair
$(h, dV)$
determine an $\lcal^2$ norm on $H^0(M, L^N)$ and hence a  Gaussian
probability
measure $\mu_N$.
All probabilistic notions such as expectations or correlations are with
reference to this measure.
The basic theme of the results of \cite{BSZ1,BSZ2} was that in a certain
scaling limit, the
correlations between zeros are universal in the sense of being independent
of $M, L, \omega$,
and other details of the setting.

The geometric setting was extended even further by two of the authors in
\cite{SZ2} by allowing $(M, \omega)$
to be any compact symplectic manifold
with integral symplectic form, i.e. $\frac{1}{\pi}[\omega] \in H^2(M, \Z)$. 
Complex line bundles with
$c_1(L) = \frac{1}{\pi}[\omega]$ are known in this context as  `pre-quantum
line bundles' (cf.\ \cite{W}). It has
been known for some time \cite{BG} that there are good analogues of
holomorphic sections
of powers of such line bundles in this context.  Interest
in symplectic analogues of holomorphic line bundles and their holomorphic
sections has grown recently because of Donaldson's \cite{DON.1} use of
asymptotically holomorphic sections of powers of pre-quantum
line bundles over symplectic manifolds in constructing embedded symplectic
submanifolds, 
Lefschetz pencils and other constructions of an algebro-geometric nature
\cite{DON.1,DON.2,A,A.2,A.3,BU.1,BU.2,Sik}. Given an almost complex
structure $J$ on $M$ which is compatible with
$\omega$, we
follow Boutet de Monvel - Guillemin \cite{BG} (see also \cite{GU} and
\cite{BU.1,BU.2}) in defining   spaces $H^0_J(M, L^N)$ of {\it almost
holomorphic} sections of the
pre-quantum line bundle $L \to M$ with curvature $\omega.$ A hermitian metric
$h$
on $L$ and $\omega$ determine an $\lcal^2$ norm and hence a Gaussian measure
$\mu_N$ on
$H^0_J(M, L^N)$.

  Our main concern
is with the zeros
$Z_s$ of $k$-tuples $s=(s_1,\dots,s_k)$ of holomorphic or
almost-holomorphic  sections.
We let $|Z_s|$ denote Riemannian $(2m-2k)$-volume on $Z_s$, regarded
as a measure on $M$: $$(|Z_s|,\phi)=\int_{Z_s}\phi d\vol_{2m-2k}\,.$$
As in \cite{BSZ1,BSZ2}, we introduce the punctured product
\begin{equation}\label{Mn}M_n=\{(z^1,\dots,z^n)\in
\underbrace{M\times\cdots\times M}_n: z^p\ne z^q \ \ {\rm for} \ p\ne
q\}\end{equation} and  consider the product measures on
$M_n$,
$$|Z_s|^n:=\big(\underbrace{|Z_s|\times\cdots\times|Z_s|}_{n}\big)\,.$$
The expectation $\E |Z_s|^n$
is called the {\it $n$-point zero correlation measure.} We write
$$\E |Z_s|^n=K_{nk}^N(z^1,\dots,z^n)dz\,,$$
where $dz$ denotes the product volume form on $M_n$. The generalized
function
$K_{nk}^N(z^1,\dots,z^n)$ is called the {\it $n$-point zero correlation
function\/}.

The main points are first to 
express these correlation measures in terms of the joint probability 
distribution  $$\wt
\D^N_{(z^1,\dots,z^n)}=\wt D^N(x^1,\dots,x^n,
\xi^1,\dots,\xi^n;z^1,\dots, z^n) dx d\xi$$ of the random variables 
$s(z^1),\dots,s(z^n),
\nabla s(z^1),\dots,\nabla s(z^n)$, and secondly to prove that the
latter has a universal scaling  limit.  Here $dx$ denotes volume measure on
$L^N_{z^1}\oplus\cdots\oplus L^N_{z^n}$, and $d\xi$ is volume measure on
$(T^*_M\otimes L^N)_{z^1}\oplus\dots\oplus(T^*_M\otimes L^N)_{z^n}$. For
more details and precise definitions, see \S \ref{s-general}.  As for the
first point, we have the following formula for the correlation
measures in terms of the joint probability distribution:

\begin{theo}\label{introdn} The $n$-point zero correlation function for
random almost holomorphic sections of $L^N\to M$ is given by
$$K_{nk}^N(z)=
\int d\xi\,\wt D_n^N(0,\xi,z)\prod_{p=1}^n
\sqrt{\det(\xi^{p}\xi^{p*})}\,.$$
\end{theo}

One of our main results is Theorem \ref{corr}, which gives a general form
of Theorem \ref{introdn} with $H^0_J(M,L^N)$ replaced by a finite
dimensional space of sections of an arbitrary vector bundle over a
Riemannian manifold. Theorem
\ref{corr} is a generalization of the Kac-Rice formula \cite{ Ka,Ri} (see
also \cite{BD,EK, Halp, ShSm}) to higher dimensions.  A special case of
Theorem
\ref{corr} was given by J.~Neuheisel \cite{Ne} in a parallel study of
correlations of nodal sets (zero sets of eigenfunctions of the Laplacian) on
spheres.

The correlations of course depend heavily on the geometry of the bundle.  
For instance,
it was shown in \cite{SZ} that  $Z_{s_N} \to \omega$ for almost every
sequence $\{s_N\}$ of  holomorphic sections
of $L^N$.  That is, zeros tend almost surely  to congregate in highly 
(positively) curved regions.
To
find universal quantities, we scale around a point $z_0 \in M$. The most
vivid
case is where $k =m$ so that almost surely the simultaneous zeros of the
$k$-tuple of sections form a discrete set.  The density of zeros in a unit
ball $B_{1}(z_0)$ around $z_0$ then grows like $ N^m$, so we rescale the
zeros
in the $1/\sqrt{N}$ ball $B_{1/\sqrt{N}}(z_0)$ by a factor of $\sqrt{N}$ to
get configurations of zeros with a constant density as $N \to \infty.$ Our
problem is whether the statistics of these configurations tend to a limit
and
whether the limit is universal.   In \cite[Th.~3.4]{BSZ2} (see also
\cite{BSZ1}), it was shown that when $L$ is a positive holomorphic line
bundle
over a complex manifold $M$, the scaled $n$-point correlation functions
$K_{nk}^N(\frac{z^1}{\sqrtn},\dots,\frac{z^1}{\sqrtn})$ converge in the high
power limit to a universal correlation function
$K_{nkm}^\infty(z^1,\dots,z^n)$  on  the punctured product $(\C^m)_n$
depending only on the dimension
$m$ of the manifold and the codimension $k$ of the zero set.  Our main
application of Theorem \ref{introdn} is that this universality law for the
scaling limits of the zero correlation functions
extends to the general symplectic case:

\begin{theo}\label{usls} Let $L$ be the pre-quantum line bundle over a
$2m$-dimensional compact integral symplectic manifold $(M,\om)$. Let $z_0\in
M$ and
choose complex local coordinates $\{z_j\}$ centered at
$z_0$ so that $\om|_{z_0}=\frac{i}{2}\sum dz_j\wedge d\bar z_j$  and $(\d
/\d z_j)|_{z_0}\in T^{1,0}M$ ($1\le j\le m$).  Let
$\scal=H_J^0(M,L^N)^k$ ($k\ge 1$), and give $\scal$ the standard Gaussian
measure $\mu$.  Then
$$\frac{1}{ N^{nk}} K_{nk}^N\left(\frac{z^1}{\sqrtn}, \dots,
\frac{z^n}{\sqrtn}\right) \to K_{nkm}^\infty(z^1,\dots,z^n)$$ (weakly in
$\dcal'((\C^m)_n)$),\
where $K_{nkm}^\infty(z^1,\dots,z^n)$ is the universal scaling limit
in
the K\"ahler setting.
\end{theo}

The proof of this result is similar to the holomorphic case \cite{BSZ2}.  
Using Theorem \ref{introdn}, we reduce the scaling limit of $K_n(z)$  to
that of  the joint probability density  $\wt\D^N_{(z^1,\dots, z^n)}$.  It was
shown in
\cite[Theorem~5.4]{SZ2} that the latter has a universal  scaling limit:
\begin{equation}\label{jpd} \wt\D^N_{(z^1/\sqrtn,\dots, z^n/\sqrtn)} 
\longrightarrow \D^\infty _{(z^1,\dots,n^n)} \,,\end{equation} where
$\D^\infty_{(z^1,\dots,z^n)}$ is a  universal Gaussian measure supported on
the holomorphic 1-jets, and $\{z_j\}$ are the complex local coordinates
of Theorem \ref{usls}.  

Let us say a few
words on the proof of (\ref{jpd}). Recall that a Gaussian measure on $\R^p$
is a measure of the form
\begin{equation}\ga_\De = \frac{e^{-\half\langle\De\inv
x,x\rangle}}{(2\pi)^{p/2}\sqrt{\det\De}}
dx_1\cdots dx_p\,,\label{gaussian}\end{equation} where
$\De$ is a positive definite symmetric $p\times p$ matrix.  Since
$\wt\D^N_{(z^1,\dots,z^n)}$ is the push-forward of a  Gaussian measure, we
have
$\wt\D^N_{(z^1,\dots,z^n)}=\ga_{\De^N}\,$ where $\De^N$ is the covariance
matrix of the random variables $(s(z^p),\nabla s(z^p))$.  The main step in
the proof in \cite{SZ2}  was to  show
that the covariance matrices $\De^N$ underlying $\wt\D^N$ tend in the 
scaling
limit to a semi-positive matrix $\De^{\infty}$.  To deal with
singular measures, we  introduced a class of  generalized
Gaussians  whose
covariance matrices are only semi-positive definite.  A
generalized Gaussian is simply a Gaussian supported on the subspace
corresponding to the positive eigenvalues of the covariance matrix.   It
followed that the scaled distributions $\wt\D^N$ tend to a generalized
Gaussian $\ga_{\De^{\infty}}$ `vanishing in the
$\bar{\partial}$-directions.'  To prove that $\De^N \to 
\De^{\infty}$, we
expressed $\De^N$ in terms of the {\it Szeg\"o kernel} $\Pi_N(x, y)$ and its 
derivatives.  The
Szeg\"o kernel is essentially the orthogonal projection from $\lcal^2(M, L^N) \to 
H^0_J(M, L^N)$. Since
it is more convenient to deal with scalar kernels than sections, we pass 
from $L \to M$ to the
associated principal $S^1$ bundle $X \to M$.  Sections $s$ of $L^N$ are then 
canonically identified with
equivariant functions $\hat{s}$ on $X$ transforming by $e^{i N \theta}$ 
under the $S^1$ action. The
space $H^0_J(M, L^N)$ then corresponds to a space $\hcal^2_N(X)$ of 
equivariant functions.  In the holomorphic
case, these functions are CR functions; i.e., they  satisfy
the tangential Cauchy-Riemann equations $\partial_b \hat{s} = 0.$ In the 
symplectic almost-complex case they
are `almost CR' functions in a sense defined by Boutet de Monvel and 
Guillemin. The scalar Szeg\"o kernels
are then the orthogonal projections $\Pi_N: \lcal^2(X) \to \hcal^2_N(X).$
The main ingredient  in the proof of (\ref{jpd}) was   the {\it scaling
asymptotics} of the Szeg\"o kernels $\Pi_N(x,y)$. In  `preferred'
local coordinates  
$(z, \theta)$ on $X$ (see \S \ref{s-scaling}), the scaling
asymptotics read:
\begin{equation} \Pi_N(z_0 + \frac{u}{\sqrt{N}}, \frac{\theta}{N}, z_0 + 
\frac{v}{\sqrt{N}}, \frac{\phi}{N})
\sim e^{i (\theta - \phi)}
e^{u \cdot \bar{v} -\half (|u|^2 + |v|^2)} \{1 + \frac{1}{\sqrt{N}}
p_{1}(u,v; z_0)
+ \cdots \}\,.\end{equation}

The universal limit correlation functions $K_{nkm}^\infty(z^1,\dots,z^n)$
are described in \cite{BSZ2} (see also \cite{BSZ1}).  They are given in
terms of the level 1 Szeg\"o kernel for the (reduced) Heisenberg group (see
\S \ref{s-examples}),
\begin{equation}\label{Heisenberg}\Pi^\H_1(z,\theta;w,\phi) =
\frac{1}{\pi^m} e^{i(\theta-\phi)+i\Im
(z\cdot \bar w)-\half |z-w|^2}= \frac{1}{\pi^m} e^{i(\theta-\phi)+z\cdot
\bar
w-\half(|z|^2+|w|^2)}\,,\end{equation} and its first and second derivatives
at
the points $(z,w)=(z^p,z^{p'})$.  Indeed, the correlation functions are
universal rational functions in $z^p_q,\bar z^p_q, e^{z^p\cdot \bar
z^{p'}}$,
and are smooth functions on $(\C^m)_n$.
We let $\wt
K_{nkm}(z^1,\dots,z^n):= (K_{1km}^\infty)^{-n}K_{nkm}(z^1,\dots,z^n)$ denote
the ``normalized" $n$-point limit correlation function, where
$K_{1km}^\infty=\frac{m!}{\pi^k(m-k)!}$ is the expected volume density of the
zero set. 
For example \cite{BSZ1,BSZ2},
\begin{equation}\label{paircor} \wt K_{21m}^\infty (z^1,z^2)=
\frac{\left[\frac{1}{2}(m^2+m)\sinh^2t+t^2\right]
\cosh t
-(m+1)t\sinh t}{m^2\sinh^3t}+\frac{(m-1)}{2m},\quad
t=\frac{|z^1-z^2|^2}{2}.\end{equation}
Formula (\ref{paircor}) with
$m=1$ agrees with the scaling limit pair correlation function of Hannay
\cite{Ha} (see also
\cite{BBL}) for zeros of polynomials in one complex variable, i.e. for
$M=\CP^1$ and
$L=\ocal(1)$.

The correlations are ``short range" in the sense that $\wt
K_{nkm}(z^1,\dots,z^n)=1+O(r^4 e^{-r^2})$, where $r$ is the minimum distance
between the points $z^p$ \cite{BSZ2}.  We show in \S \ref{decay} that in
fact the ``connected $n$-point correlations" are $o(e^{-R^2/n})$, where
$R$ is the maximum distance between the points.

\section{Line bundles on complex manifolds}\label{s-acsymp}

We begin with some notation and basic properties of sections of holomorphic
line bundles, their zero sets, and Szeg\"o kernels.  We
also provide two examples that will serve as model cases for studying
correlations of zeros of sections of line bundles in the high power limit.

\subsection{Sections of holomorphic line bundles}\label{cxgeom}

Let $L\to M$ be a holomorphic line bundle over a compact complex manifold. 
Thus, at each
$z \in M$, $L_z \simeq \C$ is a complex line and locally, over a 
sufficiently small open set $U \subset M$,
$L \simeq U \times \C.$
For background on line bundles and other objects of  complex geometry,
we refer to \cite{GH}.

A key notion is that of {\it positive} line bundle. By definition, this 
means that there exists
a smooth Hermitian metric
$h$ on $L$ with
positive curvature form
\begin{equation}\label{curvature}\Theta_h=-\ddbar
\log\|e_L\|_h^2\;,\end{equation} where $e_L$ denotes a local holomorphic
frame
(= nonvanishing section) of $L$ over an open set $U\subset M$, and
$\|e_L\|_h=h(e_L,e_L)^{1/2}$ denotes the $h$-norm of $e_L$.
A basic example is the hyperplane bundle $\mathcal O(1) \to \CP^m$, the dual 
of the
tautological line bundle. When $m = 1$, its square is the holomorphic 
tangent bundle
$T \CP^1$.  Its positivity is equivalent to the positivity of the curvature 
of $\CP^1$
in the usual sense of differential geometry. Hyperbolic surfaces ${\bf 
H}^2/\Gamma$ have
negatively curved tangent bundles, but their cotangent bundles $T^*({\bf 
H}^2/\Gamma)$
are positively curved. In the case of complex tori $\C/ \Lambda$ (where 
$\Lambda \subset \C$ is a lattice),
both the tangent and cotangent bundles are flat.  The positive  
`pre-quantum' line bundle there is the bundle
whose sections are theta functions.

Intuitively speaking,  positive curvature  at $w$ creates a potential well  
which traps a particle near $z$.
On the quantum level, this particle is a wave function (holomorphic section) 
  $\Pi_N(z,w)$ which is concentrated at $w$.
This wave function is known to mathematicians as the `Szeg\"o kernel', and 
to physicists as the `coherent state' centered
at $w$.  The simplest (but non-compact) case is where $M = \C^m$ and where
$\Theta_h= \sum_{j = 1}^m d z_j \wedge d \bar{z}_j$
(cf. \cite{DON.1}).
We note that $\Theta_h=  d A,$ where $ A = \frac{1}{2}
\sum_{j = 1}^m z_j  d \bar{z}_j - \bar{z}_j d z_j$ is
a connection form on the trivial bundle $L = \C^m \times \C \to \C^m.$ The 
associated  covariant derivative on sections is given by  $\bar{\partial}_A 
f = \bar{\partial}f
+ A^{0,1} f,$ where $A^{0,1}$ is the $(0,1)$ component of $A$.  Then 
$\bar{\partial}_A e^{-|z|^2/2} = 0,$ i.e.
there is a Gaussian holomorphic section concentrated at $w = 0.$  As will be 
seen in \S \ref{s-examples}, it is essentially the
Szeg\"o kernel of the Heisenberg group.

According to the above intuition, positive line bundles should have a 
plentiful supply of global holomorphic sections. Indeed, the  space
$H^0(M, L^{N})$ of  holomorphic sections of
$L^N=L\otimes\cdots\otimes L$ is a complex vector space of dimension
$d_N 
= \frac{c_1(L)}{m!} N^m + \cdots  $ given by the Hilbert polynomial
(\cite{Kl,Na}; see \cite[Lemma 7.6]{SS}).  It is in part  because the
dimension
$d_N$ increases so rapidly with
$N$ that probabilities  and correlations simplify so much as $N \to
\infty.$

To define the term `Szeg\"o kernel' we need to define a Hilbert space 
structure on $H^0(M, L^{N})$:
We
give $M$
the
Hermitian metric corresponding to the \kahler form
$\omega=\frac{\sqrt{-1}}{2}\Theta_h$ and the induced Riemannian volume
form
\begin{equation}\label{dV} dV_M= \frac{1}{m!}
\omega^m\;.\end{equation}
Since $\frac{1}{\pi}\om$ is a
de Rham representative of the Chern class $c_1(L)\in H^2(M,\R)$, it follows
from (\ref{dV}) that 
$\vol (M)=\frac{\pi^m}{m!}c_1(L)^m$.

The metric $h$ induces Hermitian metrics
$h^N$ on $L^N$ given by $\|s^{\otimes N}\|_{h^N}=\|s\|_h^N$.  We give
$H^0(M,L^N)$ the Hermitian inner product
\begin{equation}\label{inner}\langle s_1, s_2 \rangle = \int_M h^N(s_1,
s_2)dV_M \quad\quad (s_1, s_2 \in H^0(M,L^N)\,)\,.\end{equation}
We first define the Szeg\"o kernels as the orthogonal projections $\Pi_N:
\lcal^2(M,  L^N) \to H^0(M, L^N)$.
The projections $\Pi_N$ can be given in terms of  orthonormal bases
$\{S_j^N\}$  of sections of $H^0(M, L^N)$ by
\begin{equation}\label{szego}\Pi_N(z,w)=\sum_{j=1}^{d_N}
S_j^N(z)\otimes\overline{ S_j^N(w)}\,,\end{equation} so that
\begin{equation} (\Pi_N s)(w)  = \int_M h^N_z\big(s(z),\Pi_N(z,w)\big)
dV_M(z)\,,
\quad s\in \lcal^2(M, L^N)\,.
\end{equation}
Since we are studying the asymptotics of the  $\Pi_N$ as $N\to\infty$, we
find it useful to instead view the \szego kernels as projections on the same
space of functions.  We show how this is accomplished below.

\subsection{Lifting the Szeg\"o kernel}

As in \cite{BG, Ze,SZ, BSZ2, SZ2} and elsewhere, we analyze the $N
\to
\infty$ limit  by  lifting the analysis of holomorphic sections
over $M$ to
a certain $S^1$ bundle $X \to M$.
We let $L^*$ denote the dual line bundle to $L$, and we consider the
circle
bundle $X=\{\la \in L^* : \|\la\|_{h^*}= 1\}$, where $h^*$ is the norm
on
$L^*$ dual to $h$. Let $\pi:X\to M$ denote the bundle map; if $v\in
L_z$, then
$\|v\|_h=|(\la,v)|$, $\la\in X_z=\pi^{-1}(z)$. Note that $X$ is the boundary
of the
disc bundle $D = \{\la \in L^* : \rho(\la)>0\}$, where
$\rho(\la)=1-\|\la\|^2_{h^*}$. The disc bundle $D$ is strictly
pseudoconvex in
$L^*$, since $\Theta_h$ is positive, and hence $X$ inherits the
structure of
a strictly pseudoconvex CR manifold.  Associated to $X$ is the contact
form
$\al= -i\partial\rho|_X=i\dbar\rho|_X$.  We also give $X$ the volume
form
\begin{equation}\label{dvx}dV_X=\frac{1}{m!}\al\wedge
(d\al)^m=\al\wedge\pi^*dV_M\,.\end{equation}

The setting for our analysis of the Szeg\"o kernel is the Hardy space
$\hcal^2(X)
\subset \lcal^2(X)$ of square-integrable CR functions on $X$, i.e.,
functions that are annihilated by the
Cauchy-Riemann operator $\dbar_b$ (see \cite[pp.~592--594]{Stein}) and are
$\lcal^2$ with respect to the inner product
\begin{equation}\label{unitary} \langle  F_1, F_2\rangle
=\frac{1}{2\pi}\int_X
F_1\overline{F_2}dV_X\,,\quad F_1,F_2\in\lcal^2(X)\,.\end{equation}
Equivalently, $\hcal^2(X)$
is the space of boundary values of holomorphic functions on $D$ that
are
in
$\lcal^2(X)$.  We let $r_{\theta}x =e^{i\theta} x$ ($x\in X$) denote the
$S^1$
action on $X$ and denote its infinitesimal generator by
$\frac{\partial}{\partial \theta}$. The $S^1$ action on $X$ commutes
with $\bar{\partial}_b$; hence $\hcal^2(X) = \bigoplus_{N
=0}^{\infty} \hcal^2_N(X)$ where $\hcal^2_N(X) =
\{ F \in \hcal^2(X): F(r_{\theta}x)
= e^{i
N \theta} F(x) \}$. A section $s_N$ of $L^N$ determines an equivariant
function
$\hat{s}_N$ on $L^*$ by the rule 
$$\hat{s}_N(\lambda) = \left( \lambda^{\otimes N}, s_N(z)
\right)\,,\quad
\la\in L^*_z\,,\ z\in M\,,$$
where $\lambda^{\otimes N} = \lambda \otimes
\cdots\otimes
\lambda$. We henceforth
restrict
$\hat{s}$ to $X$ and then the equivariance property takes the form
$\hat s_N(r_\theta x) = e^{iN\theta} \hat s_N(x)$. The map $s\mapsto
\hat{s}$ is a unitary equivalence between $H^0(M, L^{ N})$ and
$\hcal^2_N(X)$. (This follows from (\ref{dvx})--(\ref{unitary}) and the fact
that
$\alpha= d\theta$ along the fibers of $\pi:X\to M$.)

We now define the (lifted) \szego kernel to be the  orthogonal projection
$\Pi_N : \lcal^2(X)\rightarrow
\hcal^2_N(X)$. It is defined by
\begin{equation} \Pi_N F(x) = \int_X \Pi_N(x,y) F(y) dV_X (y)\,,
\quad F\in\lcal^2(X)\,.
\label{PiNF}\end{equation} As above, it can be given as
\begin{equation}\label{szego2}\Pi_N(x,y)=\sum_{j=1}^{d_N}\wh
S_j^N(x)\overline{\wh S_j^N(y)}\,,\end{equation} where
$S_1^N,\dots,S_{d_N}^N$ form an orthonormal basis of $H^0(M,L^N)$.
Note that although the Szeg\"o kernel $\Pi_N$ is defined on $X$, its absolute
value is well-defined on $M$. In particular, on the diagonal we have
$$\Pi_N(z,z)=
\Pi_N(z,\theta;z,\theta)=\sum_{j=1}^{d_N}\|S_j^N(z)\|^2_{h^N}\,.$$

\subsection{Model examples}\label{s-examples}  
The Szeg\"o
kernels and their derivatives were worked out explicitly in
\cite{BSZ2} for two model cases, namely for the hyperplane
section bundle over  $\CP^m$ and for the Heisenberg bundle over
$\C^m$, i.e. the trivial line bundle with curvature equal to the
standard symplectic form on $\C^m$.
These cases are
important, since by universality, the scaling limits of
correlation functions
for all line bundles coincide with those of the  model cases.

In fact, the two models are locally  equivalent in the CR sense.
In the case of
$\CP^m$, the circle bundle $X$ is the $2m + 1$ sphere $S^{2m +
1}$, which is the boundary of the unit ball $B^{2m + 2} \subset
\C^{m+1}$. In the case of $\C^m$, the
circle bundle is the reduced Heisenberg group ${\bf H}^m_{\rm red}$, which
is a discrete quotient of the simply connected
Heisenberg group $\C^m \times \R$.

We summarize here the formulas for the \szego kernels from \cite{BSZ2} in
these model cases; for further details see \cite[\S 1.3]{BSZ2}.  For the first
example (see also \cite[\S 4.2]{SZ}), $M=\CP^m$ and $L$ is the hyperplane
section bundle $\ocal(1)$.  Sections $s\in H^0(\C\PP^m,\ocal(1))$ are linear
functions on $\C^{m+1}$, so that the zero divisors $Z_s$ are projective
hyperplanes. The line bundle $\ocal(1)$ carries a natural metric $h_\FS$ given
by
\begin{equation}\label{hfs} \|s\|_{h_\FS}([w])=\frac{|(s,w)|}{|w|}\;,
\quad\quad
w=(w_0,\dots,w_m)\in\C^{m+1}\;,\end{equation} for $s\in\C^{m+1*}\equiv
H^0(\C\PP^m,\ocal(1))$, where $|w|^2=\sum_{j=0}^m |w_j|^2$ and
$[w]\in\C\PP^m$
denotes the complex line through $w$. The \kahler form on $\CP^m$ is the
Fubini-Study form
\begin{equation}
\omega_\FS=\frac{\sqrt{-1}}{2}\Theta_{h_\FS}=\frac{\sqrt{-1}}{2}
\ddbar \log |w|^2 \,.\end{equation}
The dual bundle $L^*=\ocal(-1)$ is the affine space
$\C^{m+1}$ with the origin blown up, and $X=S^{2m+1}\subset\C^{m+1}$.
The
$N^{\rm th}$ tensor power of $\ocal(1)$ is denoted $\ocal(N)$.    An 
orthonormal basis for the space $H^0(\C\PP^m,\ocal(N))$ of homogeneous
polynomials on $\C^{m+1}$ of degree
$N$ is the set of monomials:
\begin{equation}\label{orthonormal}s^N_J =
\left[\frac{(N+m)!}{\pi^m j_0!\cdots j_m!}\right]^\half z^J\,,\quad z^J=
z_0^{j_0}\cdots z_m^{j_m},\quad\quad J=(j_0,\ldots,j_m),\
|J|=N\end{equation} Hence the
Szeg\"o kernel for $\ocal(N)$ is given by
\begin{equation}\label{szegosphere} \Pi_N(x,y)=\sum_J
\frac{(N+m)!}{\pi^mj_0!\cdots j_m!}x^J \bar y^J =  \frac{(N+m)!}
{\pi^mN!}\langle x,y\rangle^N\,.\end{equation}
Note that
$$\Pi(x,y)=\sum_{N=1}^\infty\Pi_N(x,y)=\frac{m!}{\pi^m}
(1-\langle x, y\rangle)^{-(m+1)}\,,$$ which is
the classical \szego kernel for the $(m+1)$-ball.

The second
example is  the linear model $\C^m
\times \C \to \C^m$  for  positive line bundles $L \to M$ over \kahler
manifolds and their associated Szeg\"o kernels.  Its associated
principal $S^1$ bundle $\C^m \times S^1 \to \C^m$, which may be
identified with the boundary of the disc bundle $D \subset L^*$ in the
dual line bundle, is the {\it
reduced Heisenberg group} ${\bf H}_{\rm red}^m$.
Let us summarize its definition and properties. We start  with the
usual (simply connected) Heisenberg group ${\bf
H}^m=\C^m \times \R$ with  group law
$$(\zeta, t) \cdot (\eta, s) = (\zeta + \eta, t + s +  \Im (\zeta
\cdot
\bar{\eta})).$$
The identity element is $(0, 0)$ and $(\zeta, t)^{-1} = (- \zeta, - t)$.
The  Lie algebra of ${\bf H}_m$ is spanned by elements $Z_1, 
\dots, Z_m, \bar{Z}_1, \dots, \bar{Z}_m, T$ satisfying the canonical 
commutation relations $[Z_j, \bar{Z}_k] = -i  \delta_{jk}  T$ (all other 
brackets are zero). Below
we will select such a basis of left invariant vector fields.

We can regard ${\bf H}^m$ as a strictly convex CR manifold
which may be embedded in $\C^{m + 1}$ as the boundary of a strictly
pseudoconvex domain, namely
the upper half space $ \ucal ^m :=  \{z \in \C^{m + 1}: \Im z_{m + 1}
>\half  \sum_{j = 1}^m |z_j|^2 \}$.
${\bf H}^m$ acts simply transitively on $\partial \ucal ^m$ (cf.
\cite{Stein}, XII), and we get
an identification of ${\bf H}^m$ with $\partial \ucal ^m$ by:
$$[\zeta, t] \to (\zeta, t + i |\zeta|^2) \in \partial \ucal ^m.$$

The linear model for the principal $S^1$ bundle is the reduced Heisenberg
group ${\bf H}^m_{\rm red}={\bf H}^m/ \{(0, 2\pi k): k \in \Z\} = \C^m \times
S^1$ with group law $$(\zeta, e^{i t}) \cdot (\eta, e^{i s}) = (\zeta + \eta,
e^{i[t + s + \Im (\zeta \cdot \bar{\eta})]}).$$ It is the principal $S^1$
bundle over $\C^m$ associated to the line bundle $L_\H=\C^m\times\C$.  The
metric on $L_\H$ with curvature $\Theta=\sum dz_q\wedge d\bar z_q$ is given
by setting
$h_\H(z)=e^{-|z|^2}$; i.e., $|f|_{h_\H}= |f|e^{-|z|^2/2}$.  The reduced group
${\bf H}^m_{\rm red}$ may be viewed as the boundary of the dual disc bundle $D
\subset L^*_\H$ and hence is a strictly pseudoconvex CR manifold.

We then define the Hardy space
$\mathcal{H}^2 ({\bf H}^m_{\rm red})$ of CR holomorphic
functions to be the functions in $\lcal^2({\bf H}^m_{\rm red})$ satisfying
the left-invariant Cauchy-Riemann equations 
$\bar{Z}^L_q f = 0$ ($1\le q\le m$) on ${\bf H}^m_{\rm red}$.
Here, $\{\bar{Z}^L_q\}$ denotes a basis of the left-invariant 
anti-holomorphic vector fields on ${\bf H}^m_{\rm red}$.
Let us recall their definition: we first  equip ${\bf H}^m_{\rm red}$  with 
its  left-invariant  connection form $\alpha^L = \half (
\sum_q(u_qdv_q-v_qdu_q) +  d\theta)$
($\zeta=u+iv$),
whose curvature equals  the symplectic form
$\omega = \sum_q du_q \wedge dv_q$.  The left-invariant (CR-) holomorphic 
(respectively anti-holomorphic) vector fields $Z_q^L$
(respectively $ \bar{Z}_q^L$) are
the  horizontal lifts  of the vector fields $\frac{\partial}{\partial z_q},$ 
  respectively
$\frac{\partial}{\partial \bar z_q}$ with respect to $\alpha^L$.  They span
the  left-invariant CR structure
of ${\bf H}^m_{\rm red}$ and are given by
$$Z_q^L =  \frac{\partial}{\partial
z_q} + \frac{i}{2}\bar{z}_q \frac{\partial}{\partial \theta},
\;\;\;\; \bar{Z}^L_q =
\frac{\partial}{\partial \bar{z}_q} - \frac{i}{2} z_q
\frac{\partial}{\partial
\theta}.$$
The vector fields $\{\frac{\partial}{\partial
\theta}, Z_q^L, \bar{Z}_q^L\}$ span the Lie algebra of ${\bf H}^m_{\rm red}$ 
and satisfy
the canonical  commutation relations above.

For $N=1,2,\dots$, we define  $\hcal^2_N\subset \hcal^2(\H^m_{\rm red})$ as
the (infinite-dimensional) Hilbert space of square-integrable CR functions
$f$ such that
$f\circ r_\theta = e^{iN\theta}f$ as before.  The Szeg\"o kernel $\Pi_N^\H(x,y)$ is 
the orthogonal projection to  $\hcal^2_N$.  It is given by 
\begin{equation}\label{szegoheisenberg} \Pi_N^\H(x,y)  =\frac{1}{\pi^m} N^m 
e^{i N (t-s )} e^{ N(\zeta\cdot\bar \eta -\half |\zeta|^2 -\half|\eta|^2) }
\,,\qquad x=(\zeta,t)\,,\ y=(\eta,s)\,.\end{equation}

The \szego kernels $\Pi_N^\H$ are Heisenberg dilates of the level 1 kernel
$\Pi_1^\H$: \begin{equation}\Pi^\H_N(x, y) = N^m \Pi^\H_1 (\de_{\sqrtn}\,
x,
\de_{\sqrtn}\, y)\,,\end{equation}
where the Heisenberg dilations (or scalings) $\de_r$ are the  automorphisms
of ${\bf H}^m$ $$ \delta_r (z,\theta) = (r z, r^2 \theta)\,,\quad r
\in \R^+\,.$$ (The dilation $\de_{\sqrtn}$ descends to a homomorphism of
 ${\bf H}^m_{\rm red}$.)

\begin{rem}
The group ${\bf H}^m_{\rm red}$ acts by left translation on 
$\mathcal{H}^2_1$.  The generators of this
representation are the right-invariant vector fields $Z_q^R, \bar{Z}_q^R$ 
together with $\frac{\partial}{\partial
\theta}$. They are horizontal
with respect to the right-invariant contact form $\alpha^R = \half ( 
\sum_q(u_qdv_q-v_qdu_q) - d\theta)$ and are given by:
$$Z_q^R =  \frac{\partial}{\partial
z_q} - \frac{i}{2}\bar{z}_q \frac{\partial}{\partial \theta},
\;\;\;\; \bar{Z}^R_q =
\frac{\partial}{\partial \bar{z}_q} + \frac{i}{2} z_q
\frac{\partial}{\partial \theta}.$$
In physics terminology, $Z_q^R$ is known as an annihilation operator and 
$\bar{Z}_q^R$ is a creation operator.

The representation $\mathcal{H}^2_1$ is irreducible and   may be identified 
with the Bargmann-Fock space of entire holomorphic functions on $\C^n$ which
are square integrable relative to $e^{-|z|^2}$.  The identification goes as 
follows:
the function    $\phi_o(z, \theta) :=  e^{i \theta} e^{-|z|^2/2}$ is 
CR-holomorphic and is  also the ground state for the  right invariant 
annihilation operator; i.e., it  satisfies $$\bar{Z}^L_q
\phi_o(z, \theta) = 0 = Z_q^R \phi_o(z, \theta).$$
In the physics terminology, the level 1  Szeg\"o kernel $\Pi_1^\H$, which
is  the left translate of  $\phi_o$ by $(-w, - \phi)$, 
is the coherent state associated to
the phase space point $w$. Any  element 
$F(z,\theta)$ of $\mathcal{ H}^2_1$ may be written in the form $F (z, \theta) 
= f(z) \phi_o.$  Then   $\bar{Z}^L_q F =(\frac{\partial}{\partial \bar{z}_q} 
f ) \phi_o$, so that $F$ is CR if and only if $f$
is holomorphic. Moreover, $F \in \lcal^2({\bf H}^m_{\rm red})$ if and only 
if $f$ is square
integrable relative to $e^{-|z|^2}$.

\end{rem}

\section{Almost-complex symplectic manifolds}  In \cite{SZ2}, the study
of the \szego kernel was extended to almost-complex symplectic manifolds,
and parametrices and resulting off-diagonal asymptotics for the \szego
kernel were obtained in this general setting.  We now summarize the basic
geometric and analytic constructions of  \cite{SZ2} for the almost-complex
symplectic case.

We denote by $(M, \omega)$ a compact 
symplectic manifold such that $[\frac{1}{\pi}\omega]$ is an integral
cohomology class. We also fix a compatible almost complex structure $J$
satisfying $\omega(v, Jv) > 0$.  We denote by $T^{1,0}M, $ respectively $T^{0, 1}M$,
the holomorphic (respectively  anti-holomorphic) sub-bundles of the complex tangent
bundle, i.e. $J = i$ on $T^{1,0}$ and $J = -i$ on $T^{0,1}$.  It is well known
(see \cite[Prop.~8.3.1]{W}) that there exists a Hermitian line bundle $(L,
h)
\to M$ and a metric connection $\nabla$ on $L$ whose curvature $\Theta_L$
satisfies $\frac{i}{2} \Theta_L = \omega$.  The `quantization' of $(M,
\omega)$ at Planck constant $1/N$ should be a Hilbert space of polarized
sections of the $N^{\rm th}$ tensor power $L^N$ of $L$ (\cite[p.~266]{GS}). In
the complex case, polarized sections are simply holomorphic sections.  The
notion of polarized sections is problematic in the non-complex symplectic
setting, since the Lagrangean subspaces $T^{1,0}M$ defining the complex
polarization are not integrable and there usually are no `holomorphic'
sections.  A subtle but compelling replacement for the notion of polarized
section has been proposed by Boutet de Monvel and Guillemin \cite{BG}, and it
is this notion which was used in \cite{SZ2}.

To define these polarized sections, we work as above on the associated
principal $S^1$ bundle $X \to M$ with $X = \{v \in L^*: |v|_h =1\}$.  We let
$\alpha$ be the connection 1-form on $X$ given by $\nabla$; we then have
$\frac{1}{\pi}d\alpha =\pi^* \omega$, and thus $\al$ is a contact form on $X$,
i.e., $\al\wedge (d\al)^m$ is a volume form on $X$.  In the complex case, $X$
was a CR manifold.  In the general almost-complex symplectic case it is an
almost CR manifold.  The almost CR structure is defined as follows: The kernel
of $\alpha$ defines a horizontal hyperplane bundle $H \subset TX$. Using the
projection $\pi: X \to M$, we may pull back $J$ to an almost complex structure
on $H$. We denote by $H^{1,0}$, respectively $H^{0,1}$ the eigenspaces of
eigenvalue $i$, respectively $-i$, of $J$.  The splitting $TX = H^{1,0} \oplus
H^{0,1} \oplus \C \frac{\partial}{\partial \theta}$ defines the almost CR
structure on $TX$.  We also define local orthonormal frames $Z_1, \dots, Z_n$
of $H^{1,0}$, respectively $\bar{Z}_1, \dots, \bar{Z}_m$ of $H^{0,1}$, and
dual orthonormal coframes $\vartheta_1, \dots, \vartheta_m,$ respectively
$\bar{\vartheta}_1, \dots, \bar{\vartheta}_m$. On the manifold $X$ we have $d=
\d_b +\dbar_b +\frac{\partial}{\partial \theta}\otimes \alpha$, where
$\partial_b = \sum_{j = 1}^m {\vartheta}_j \otimes{Z}_j$ and $\dbar_b =
\sum_{j = 1}^m \bar{\vartheta}_j \otimes \bar{Z}_j$.  Note that for an
$\lcal^2$ section $s^N$ of $L^N$, we have
\begin{equation*}
(\nabla_{L^N}s^N)\nhat = d^h\hat s^N\,,\end{equation*} where
$d^h=\d_b+\dbar_b$ is the horizontal derivative on $X$.

\subsection{The $\bar{D}$ complex and Szeg\"o kernels}\label{s-dbarcomplex}

In the complex case, a holomorphic section $s$ of $L^N$ lifts to a function
$\hat{s}\in \lcal^2_N(X)$ satisfying $\dbar_b \hat{s} = 0$. The
operator $\dbar_b$ extends to a complex satisfying $\dbar_b^2 = 0$,
which is a necessary
and sufficient condition for having a maximal family of CR holomorphic
coordinates.
In the non-integrable case  $\dbar_b^2 \not= 0$, and there may be no
solutions of $\dbar_b f = 0.$  To define polarized sections and their
equivariant lifts,
Boutet de Monvel \cite{Bou} and Boutet de Monvel - Guillemin \cite{BG}
defined a complex $\bar{D}_j$, which is a good replacement for
$\dbar_b$ in the non-integrable
case.  Their main result is:

\begin{theo}\label{COMPLEX} {\rm \cite{BG}, Lemma 14.11 and Theorem A
5.9)} There exists  an $S^1$-invariant  complex of first order
pseudodifferential operators $\bar{D}_j$ over $X$
$$0 \rightarrow C^{\infty}(\Lambda_b^{0,0})
\ {\buildrel \bar{D}_0 \over\to}\  C^{\infty}(\Lambda_b^{0,1})
\ {\buildrel \bar{D}_1 \over\to}\  \cdots \ {\buildrel \bar{D}_{m-1}
\over\longrightarrow}\
C^{\infty}(\Lambda_b^{0,m})\to 0\,,$$ where $\Lambda_b^{0,j}=\La^j
(H^{0,1}X)^*$, such that:

\begin{enumerate}
\item[{\rm i)}]  $\sigma(\bar{D}_j) =
\sigma(\bar{\partial}_b)$ to second order along $\Sigma:=\{(x,r\al_x):x\in X,
r>0\}\subset T^*X$;
\item[{\rm ii)}] The orthogonal projector  $\Pi : \lcal^2(X)
\to
\hcal^2(X)$ onto the
kernel of $\bar{D}_0$ is a complex Fourier integral operator
which is microlocally equivalent to the Cauchy-Szeg\"o projector of the
holomorphic case; 
\item[{\rm iii)}] $(\bar{D}_0,
\frac{\partial}{\partial \theta})$ is jointly elliptic. 
\end{enumerate}\end{theo}

We refer to the kernel $\hcal^2(X)=\ker \bar D_0 \cap \lcal^2(X)$ as the Hardy
space of square-integrable `almost CR functions' on $X$. The $\lcal^2$ norm is
with respect to the inner product (\ref{unitary}) as in the holomorphic case.
Since the $S^1$ action on $X$ commutes with $\bar{D}_0$, we have as before the
decomposition $\hcal^2(X) = \bigoplus_{N =0}^{\infty} \hcal^2_N(X)$, where
$\hcal^2_N(X)$ denotes the almost CR functions on $X$ that transform by the
factor $e^{i N \theta}$ under the action $r_\theta$. By property (iii)
above, they are smooth functions.  We denote by $H^0_J(M, L^{ N})$ the space
of sections corresponding to $\hcal^2_N(X)$ under the map $s\mapsto
\hat{s}$.  Elements of
$H^0_J(M, L^{ N})$ are the `almost holomorphic sections' of $L^N$. (Note
that products of almost holomorphic sections are not necessarily almost
holomorphic.) We henceforth identify $H^0_J(M, L^{ N})$ with
$\hcal^2_N(X)$.  By the Riemann-Roch formula of \cite[Lemma~14.14]{BG}, the
dimension of $H^0_J(M, L^N)$ (or $ \hcal^2_N(X)$) is given by $d_N =
\frac{c_1(L)}{m!} N^m + \cdots$ (for $N$ sufficiently large), as before. 
(The estimate $d_N
\sim\frac{c_1(L)}{m!} N^m$ also follows from \cite[\S 4.2]{SZ2}.)

As before, we let $\Pi_N : \lcal^2(X) \rightarrow \hcal^2_N(X)$ denote the
orthogonal
projection.  The  level $N$ Szeg\"o kernel $\Pi_N(x,y)$ is given as
in the holomorphic case  by (\ref{PiNF}) or (\ref{szego2}), using an
orthonormal basis
$S_1^N,\dots,S_{d_N}^N$ of
$H^0_J(N,L^N)\equiv \hcal^2_N(X)$.

\subsection{Scaling limit of the Szeg\"o kernel}\label{s-scaling}

Our analysis is based on the near-diagonal scaling asymptotics of
the
\szego kernel from \cite{SZ2}. These asymptotics are given in
terms of the Heisenberg dilations $\de_{\sqrtn}$, using local `Heisenberg
coordinates' at a point $x_0\in X$.  These coordinates are
given in terms of {\it preferred coordinates\/} at 
$P_0=\pi(x_0)$ and a {\it preferred frame\/} at $P_0$. A
coordinate system $(z_1,\dots,z_m)$ on a neighborhood
$U$ of
$P_0$ is said to be preferred if
$$(g-i\om)|_{P_0}=\sum_{j=1}^m d z_j\otimes d\bar z_j\big|_0
\,.$$  Here $g$ denotes the Riemannian metric
$g(v,w):=\om(v,Jw)$ induced by the symplectic form $\om$.
Preferred coordinates satisfy the following three
(redundant) conditions:
\begin{enumerate}
\item[i)] $\quad\d/\d z_j|_{P_0}\in T^{1,0}(M)$, for $1\le j\le m$,
\item[ii)] $\quad\om({P_0})=\om_0$,
\item[iii)] $\quad g({P_0} )= g_0$,
\end{enumerate}
where $\omega_0$ is the standard
symplectic form and $g_0$ is the Euclidean metric:
$$\om_0=\frac{i}{2}\sum_{j=1}^m dz_j\wedge d\bar z_j =\sum_{j=1}^m
(dx_j\otimes dy_j - dy_j\otimes dx_j)\,,\quad g_0=
\sum_{j=1}^m
(dx_j\otimes dx_j + dy_j\otimes dy_j)\,.$$
 A {\it preferred frame\/} for $L\to M$ at  $P_0$  is a local
frame ($=$nonvanishing section) $e_L$ on $U$ such that 

\begin{enumerate}
\item[i)] $\quad \|e_L\|_{P_0} =1$;
\item[ii)] $\quad \nabla e_L|_{P_0} = 0$;
\item[iii)] $\quad \nabla^2 e_L|_{P_0} = -(g+i\om)\otimes
e_L|_{P_0}\in T^*_M\otimes T^*_M\otimes L$.
\end{enumerate}
A preferred frame can be constructed by multiplying an arbitrary
frame by a function with specified 2-jet at $P_0$; any two such
frames necessarily agree to third order at $P_0$. 

\begin{defin} A  {\it Heisenberg coordinate chart\/} at a point
$x_0$ in the principal bundle $X$ is a coordinate chart
$\rho:U\approx V$ with $0\in U\subset \C^m\times \R$ and
$\rho(0)=x_0\in V\subset X$ of the form
\begin{equation}\rho(z_1,\dots,z_m,\theta)= e^{i\theta}
h(z)^{\half} e^*_L(z)\,,\label{coordinates}\end{equation} where
$e_L$ is a preferred local frame for $L\to M$ at $P_0=\pi(x_0)$,
and
$(z_1,\dots,z_m)$ are preferred coordinates centered at $P_0$.
We require that $P_0$ have coordinates $(0,\dots,0)$ and
$e_L^*(P_0)=x_0$.
\end{defin}

The following near-diagonal asymptotics of the \szego kernel
is the
 key analytical result on which our analysis of the scaling limit for
correlations of zeros  is based.

\begin{theo} \label{neardiag} {\rm (\cite{SZ2}, Theorem 2.3)}
Let $P_0\in M$ and choose a Heisenberg coordinate chart about $P_0$.
Then
$$\begin{array}{l} N^{-m}\Pi_N(\frac{u}{\sqrtn},\frac{\theta}{N};
\frac{v}{\sqrtn},\frac{\phi}{N})\\ \\ \qquad
= \Pi^\H_1(u,\theta;v,\phi)\left[1+ \sum_{r = 1}^{K}
N^{-r/2} b_{r}(P_0,u,v)
+ N^{-(K +1)/2} R_K(P_0,u,v,N)\right]\;,\end{array}$$
where $\|R_K(z_0,u,v,N)\|_{\ccal^j(\{|u|\le \rho,\ |v|\le \rho\}}\le
C_{K,j,\rho}$ for $j\ge 0,\,\rho>0$ and $C_{K,j,\rho}$ is independent of the
point
$z_0$ and choice of coordinates.
\end{theo}

This asymptotic formula has several applications to symplectic geometry, in
addition to our result on zero correlations. For example, Theorem \ref
{neardiag} is used in
\cite{SZ2} to obtain symplectic versions of
the following results in complex geometry:

\begin{itemize} \item the asymptotic expansion theorem of \cite{Ze},
\item the Tian
almost isometry theorem \cite{Ti},
\item the Kodaira
embedding theorem (see \cite{GH} or
\cite{SS}). \end{itemize} The symplectic forms of these theorems are based on
the symplectic  Kodaira maps
$\Phi_N : M \to PH^0_J(M,L^N)^*$, which are defined as in the
holomorphic case by
$\Phi_N(z) =
\{s^N: s^N(z) = 0\}$. Equivalently, we  choose an orthonormal basis
$S^N_1,\dots,S^N_{d_N}$
of $H^0_J(M,L^N)$ and write
\begin{equation*} \Phi_N : M \to\CP^{d_N-1}\,,\qquad
\Phi_N(z)=\big(S^N_1(z):\dots:S^N_{d_N}(z)\big)\,.\end{equation*}
We now state the symplectic generalizations
of the above three theorems:

\begin{theo}\label{tyz} {\rm (\cite{SZ2}, Theorems 3.1--3.2)} Let  $L
\to (M,
\omega)$ be the pre-quantum line bundle
over a $2m$-dimensional
symplectic manifold, and let $\{\Phi_N\}$ be its Kodaira maps.  Then:
\begin{itemize}

\item  There exists a complete
asymptotic expansion:
$$ \Pi_N(z,z)  =  a_0 N^m +
a_1(z) N^{m-1} + a_2(z) N^{m-2} + \dots$$
for certain smooth coefficients $a_j(z)$ with $a_0 = \pi^{-m}$.
Hence, the  maps $\Phi_N$  are well-defined for $N\gg 0$.

\item 
Let $\omega_{FS}$ denote the Fubini-Study form on $\CP^{d_N-1}$.
Then $$\|\frac{1}{N}  \Phi_N^*(\omega_{FS}) -
\omega\|_{\ccal^k} = O(\frac{1}{N})$$ for any $k$.

\item   For $N$ sufficiently large,
$\Phi_N$ is an embedding.

\end{itemize}

\end{theo}

For proofs we
refer to \cite{SZ2}. (See also \cite{BU.2} for a proof of a similar
Kodaira embedding theorem.)

\section{Correlations of zeros}\label{s-zcor}
In \S \ref{s-universality}, we shall use Theorem \ref{neardiag} and the
methods of
\cite{BSZ2} to extend the results of
\cite{BSZ1,BSZ2} on the universality of the scaling limit of the
$n$-point zero correlations to the case of almost complex symplectic
manifolds.   The basis for our argument is Theorem 2.2 from
\cite{BSZ2}, which generalizes a formula of Kac \cite{Ka} and Rice
\cite{Ri}  for zeros of functions on $\R^1$, and of \cite{Halp} for
zeros of (real) Gaussian vector fields (see also
\cite {BD,EK,Ne,ShSm}).  However, we shall need to consider the case where
the joint probability distributions are singular, and hence we give below a
complete proof of a more general result (Theorem
\ref{corr})  on the correlations of zeros of sections of
$\ccal^\infty$ vector bundles.

\subsection {General formula for zero correlations}\label{s-general}

For our general setting, we let
$(V,h)$ be a
$\ccal^\infty$ (real) vector bundle over an oriented $\ccal^\infty$
Riemannian manifold
$(M,g)$. (Here,
$h$
denotes a $\ccal^\infty$ metric on $V$.)  Suppose that $\scal$ is a finite
dimensional subspace of the space $\ccal^\infty(M,V)$ of global
$\ccal^\infty$
sections of $V$, and let $d\mu$ be a probability measure on $\scal$ given by
a
semi-positive $\ccal^0$ `rapidly decaying' volume form.  We say that a
$\ccal^0$ volume form
$\psi dx_1\wedge\cdots dx_d$ on $\R^d$ is {\it rapidly
decaying\/} if $\psi(x)=o(\|x\|^{-N})$ for all $N\in\Z^+$.  (In this
paper, we are primarily interested in the case where $d\mu$ is a Gaussian
measure.)  The purpose of
this section is to study the zero set $Z_s$ of a random section $s\in\scal$
and to obtain formulas for the expected value and $n$-point correlations of
the volume measure $|Z_s|$.  We shall later apply our results to the case
where $V=L^N\oplus\cdots\oplus L^N$, for a complex line bundle $L$ over a
compact almost complex symplectic manifold $M$ and where
$\scal=\hcal^2_N\oplus\cdots\oplus \hcal^2_N$. (Recall that
$\hcal^2_N$ is the
space of almost holomorphic sections of $L^N$.)  Then the zero sets
$Z_s$ are
the simultaneous zeros of (random) $k$-tuples of almost holomorphic
sections.

Our formulation involving
general vector bundles also allows us to reduce the study of $n$-point
correlations to the case $n=1$, i.e., to expected densities (or
volumes) of zero sets. We first describe the formula (Theorem
\ref{density}) for this expected zero density. This formula is given in
terms of the `joint probability density,' which is a measure on the
space
$J^1(M,V)$  of 1-jets of sections of
$V$. 

Recall that we have the exact sequence of vector bundles
\begin{equation}\label{jet} 0\to T^*_M\otimes V   \stackrel{\iota}{\to}
J^1(M,V) \stackrel{\pi_V}{\to}
V\to 0\,.\end{equation} We let
$$\ecal:M\times\scal\to V\,,\quad \ecal(z,s)=s(z)$$ denote the
evaluation map, and we say that $\scal$ {\it spans\/} $V$ if $\ecal$ is
surjective, i.e., if
$\{s(z):s\in\scal\}$ spans $V_z$ for all $z\in M$.  We are mainly
interested in the jet map
$$\jcal:M \times\scal\to J^1(M,V)\,,\quad \jcal (z,s)=J^1_zs=\ \mbox{the
1-jet
of}\ s\ \mbox{at}\ z\,.$$  Note that $\ecal=\pi_V\circ\jcal$.

Note that a {\it measure\/} on an $N$-dimensional manifold $Y$ is a
current
$\nu\in\dcal^0(Y)'=\dcal'{}^N(Y)$ of order 0. We can write $\nu=fd\vol_Y$,
where
$f\in
\dcal'{}^0(Y)$. (Recall that $\dcal^p(Y)$ denotes the space of compactly
supported $\ccal^\infty$ $p$-forms on $Y$, and
$\dcal'{}^p(Y)=\dcal^{N-p}(Y)'$.) Some authors refer to
$f$ as a measure, but to keep the distinction, we shall call elements of
$\dcal'{}^0(Y)$ {\it generalized functions\/}.

To describe the induced volume forms on the total spaces of the bundles in
(\ref{jet}), we write $g(z)=\sum g_{qq'}(z)du_q\otimes
du_{q'},h_{jj'}=h(e_j,e_{j'})$, where $\{u_1,\dots,u_m \}$ are local
coordinates in $M$ and $\{e_1,\dots,e_k \}$ is a local frame in $V$ ($m=\dim
M,\ k={\rm rank}\,V$). We let $G=\det (g_{qq'})$, $H=\det (h_{jj'})$.  We
further let
$dz=\sqrt{G}du_1\wedge\cdots\wedge du_m$ denote Riemannian volume in $M$,
and
we write
\begin{equation*} x=\sum_j x_j e_j(z)\in
V_z,\quad dx=\sqrt{H(z)}
dx_1\wedge\cdots\wedge dx_k\,,\end{equation*}
$$\xi=\sum_{j,q} \xi_{jq} du_q\otimes e_j|_{z}\in (T^*_M\otimes V)_{z},
\quad d\xi= G(z)^{-k/2}H(z)^{m/2} \prod_{j,q} d\xi_{jq}\,.$$
The induced
volume measures on $V$ and $T^*_M\otimes V$ are given by $dx dz$ and
$ d\xi dz$ respectively.  We give $V$ a connection that preserves $h$; its
covariant derivative provides a splitting $\nabla:J^1(M,V)\to T^*_M\otimes
V$
of (\ref{jet}), and hence $dx d\xi dz$ provides a volume form on
$J^1(M,V)$.

\begin{defin} The {\it $1$-jet density\/} of $\mu$ is the
measure
$$\D:=\jcal_* (dz\times\mu)$$ on the space $J^1(M,V)$ of 1-jets. We write
$$\D=D(x,\xi,z) dxd\xi dz\,\quad D(x,\xi,z)\in
\dcal'{}^0(J^1(M,V))\,.$$
\end{defin}

We let $\rho_\ep$ denote a $\ccal^\infty$ `approximate identity' on
$V$ of the form
$$\rho_\ep(v)= \ep^{-k}\rho(\ep\inv v)\,,
\qquad \rho\in\ccal^\infty(V)\,,\
\int_{V_z}\rho(x,z)dx=1\,,\ \rho\ge 0\,,\
\rho(v)=0\ \mbox{for\ }\|v\|\ge 1\,.$$
We let $\tilde \rho_\ep\in \ccal^\infty(J^1(M,V))$ be given by
$$\tilde \rho_\ep(x,\xi,z)=\rho_\ep(x,z)\,.$$ or formally, $\tilde \rho_\ep
=\rho_\ep\circ\pi_V$.

\begin{lem} Suppose that $\ecal$ spans $V$.  Then there exists a unique
positive measure $\D^0$ on  $T^*_M\otimes V$ such that
$$\iota_*\D^0 = \lim_{\ep\to 0} \tilde\rho_\ep \D\,.$$
Moreover, $\D^0$ is independent of the choice of local frame
$\{e_j\}$, connection $\nabla$, and approximate identity $\rho_\ep$.
\end{lem}

\begin{proof} The surjectivity of $\ecal=\pi_V\circ\jcal$ guarantees
that the normal bundle $N_\iota$ is disjoint from the wave front set of
$D(x,\xi,z)$ and hence
$\iota^*D(x,\xi,z)$ is well-defined (see \cite[Th.~8.2.4]{H}). Thus we can
define
\begin{equation}\label{D0}{\mathbf D}^0:=\iota^*D(x,\xi,z) d\xi
dz\,.\end{equation} To verify the equation of
the lemma, it suffices by the continuity of
$\iota^*$ to consider the case where $D(x,\xi,z)\in\ccal^\infty$.  In this
case, ${\mathbf D}^0=D(0,\xi,z)d\xi dz$, and hence
$$\tilde\rho_\ep {\mathbf D} \to \de_0(x)D(0,\xi,z) dxd\xi dz =
\iota_*\big(D(0,\xi,z)d\xi dz\big)=\iota_*{\mathbf D}^0\,.$$
Since $dx$ and $d\xi$ are intrinsic volume forms, it follows that
${\mathbf D}^0$ is independent of the choice of local frame
$\{e_j\}$ (and local coordinates).  To show that
$D(0,\xi,z)$ does not depend on the choice of connection on $V$, write
$s=\sum x_j e_j$, $\nabla s = \sum
\xi_{jq} dz_q \otimes e_j$, $\xi_{jq}=\frac{\d x_j}{\d z_q} + \sum_k
x_k\theta^k_{jq}$. Then if we consider the flat connection $\nabla' s
=\sum
\xi_{jq}'dz_q \otimes e_j$, $\xi_{jq}'=\frac{\d x_j}{\d z_q}$, we have
$$\frac{\d(\xi_{jq},x_j)}{\d(\xi_{jq}',x_j)} = 1\,.$$
Hence, $dxd\xi'=dx d\xi$ so that $D'(0,\xi,z)=D(0,\xi,z)$.
\end{proof}

We note that
$$(J^1_{z_0})_*\mu =D(x,\xi,z_0)dxd\xi\,,$$ so that $D(x,\xi,z_0)dxd\xi$ is
the joint probability distribution  of the random
variables $X_j^{z_0},\ \Xi_{jq}^{z_0}$ on $\scal$ given by:
$$X_j^{z_0}(s)=x_j(z_0)\,,\quad
\Xi_{jq}^{z_0}(s)=\xi_{jq}(z_0)\qquad (1\le j\le k, 1\le q
\le m)\,.$$ This is a special case  of the {\it $n$-point joint
probability distribution} defined below.

\medskip
For a vector-valued 1-form $\xi\in  T^*_{M,z}\otimes V_{z}= {\rm
Hom}(T_{M,z},V_z)$, we let $\xi^*\in {\rm Hom}(V_{z},T_{M,z})$ denote
the adjoint to $\xi$ (i.e., $\langle\xi^*v,t \rangle=\langle v,\xi
t\rangle\,$). We consider the
endomorphism
$\xi\xi^*\in {\rm Hom}(V_{z},V_{z})$, and we write
$$\bbb\xi\bbb=\sqrt{\det(\xi\xi^*)}\,.$$
(Note that $\bbb\cdot\bbb$ is not a norm.)
In terms of a local frame $\{e_j\}$,

\begin{equation}\label{localframe}
\bbb\xi\bbb=\sqrt{H}\|\xi_1\wedge\dots\wedge
\xi_k\|\,,\quad \xi=\sum_j \xi_j \otimes e_j\,.\end{equation} To verify
(\ref{localframe}), write
$$\xi_j
=\sum_{q=1}^m \xi_{jq} du_q\,;$$ then
$$\xi^*=\sum_{j,q}\xi^*_{jq}\frac{\d}{\d u_q}\otimes e_j^*\,,\qquad
\xi^*_{jq}=\sum_{j',q'} h_{jj'}\ga_{q'q}\xi_{j'q'}\,,$$
where $\big(\ga_{qq'}\big)=\big(g_{qq'}\big)^{-1}$; hence we have
\begin{equation}\label{endo} \xi\xi^*=\sum_{j,j',j'',q,q'}
h_{j'j''}\xi_{jq}\ga_{q'q} \xi_{j''q'}\, e_j\otimes e_{j'}^*
\,.\end{equation} Its determinant is given by
\begin{equation}\label{detendo}\det(\xi\xi^*)=H\det\left(\sum_{q,q'}
\xi_{jq}\ga_{q'q} \xi_{j'q'}\right)_{1\le j,j'\le k}=H\det \langle\xi_j,
\xi_{j'}\rangle=
H\|\xi_1\wedge\dots\wedge \xi_k\|^2\,,\end{equation}
which gives (\ref{localframe}).

Let us assume that $\scal$ spans $V$.  Then the incidence set
$I:=\{(z,s)\in M\times\scal: s(z)=0\}$ is a smooth submanifold and hence by
Sard's theorem applied to the projection $I\to\scal$, the zero set
$$Z_s=\{z\in M:z(s)=0\}$$ is a smooth $(m-k)$-dimensional submanifold of $M$
for almost all $s$.  (In the holomorphic case, this is called `Bertini's
Theorem.') We let $|Z_s|$ denote Riemannian $(m-k)$-volume on $Z_s$,
regarded
as a measure on $M$: $$(|Z_s|,\phi)=\int_{Z_s}\phi d\vol_{m-k} \qquad
\mbox{for a.a. \ }s\in\scal\,.$$  Its expected value is the positive
measure $\E |Z_s|$ given by $$(\E |Z_s|,\phi) =\E( |Z_s|,\phi) =
\int_\scal d\mu(s)\int_{Z_s}\phi d\vol_{m-k}\le +\infty\quad
(\phi\in\ccal^0(M)\,,\
\phi\ge 0)\,.$$ (Recall that $\E$ denotes expectation.) In fact the following
general density formula tells us that $(\E |Z_s|,\phi)<+\infty$ if the test
function
$\phi$ has compact support.

\begin{theo}\label{density} Let $M,V,\scal,d\mu$ be as above, and
suppose
that $\scal$ spans $V$.  Then
\begin{equation}\label{d1} \E|Z_s| = \pi_*(\sqrt{\det(\xi\xi^*)}\D^0)\in
\dcal'{}^m(M)
\,,\end{equation} where $\pi:T^*_M\otimes V\to M$ is the projection.
\end{theo}

Note that although $\D^0$ depends on the metric $h$ on $V$,
the measure $\sqrt{\det(\xi\xi^*)}\D^0$ is independent of $h$.  In the case
where
$D(x,\xi,z)\in\ccal^0$, (\ref{d1}) becomes
\begin{equation}\label{d2} \E|Z_s| =K_1(z)dz \,,\quad
K_1(z)=
\int D(0,\xi,z)
\sqrt{\det(\xi\xi^*)}, d\xi\,.
\end{equation} 

Before proceeding further, we first give a heuristic explanation of
(\ref{d2}). Suppose that $D\in\ccal^0$ and fix a point $z_0\in M$. Let us
consider the case where ${\rm rank}\, V=\dim M = m$ so that the zeros are
discrete. Then the probability
of finding a zero in a small ball
$\B_r=\B_r(z_0)$ of radius $r$ about $z_0$ is approximately
$K_1(z_0)\vol(\B_r)$. 
If the radius
$r$ is very small, we can suppose that the sections $s\in\scal$ are
approximately linear:
\begin{equation}\label{heuristic} s(z)\approx X^{z_0}+\Xi^{z_0}
\cdot (z-z_0)\,,\end{equation} where we have written
$s$ in terms of a local frame for $V$ and local coordinates in
$M$.   Here, $X^{z_0}=X^{z_0}(s)=\big(X^{z_0}_j(s)\big)$, respectively
$\Xi^{z_0}=\Xi^{z_0}(s)=\big(\Xi_{jq}^{z_0}(s)\big)$, is a
vector-valued, respectively matrix-valued, random variable on $\scal$.  Then
the probability that the linearized section $s$ given by
(\ref{heuristic})  has a zero in
$\B_r$ is given by
$$\mu\big\{s\in \scal:X^{z_0}\in \Xi^{z_0}(\B_r)\big\} =
\int_{\R^{m^2}} \int_{\xi(\B_r)}D(x,\xi,z_0) dx d\xi \approx
\int\vol(\xi(\B_r)) D(0,\xi,z_0) d\xi \,.$$
Since $ \vol(\xi(\B_r))= \bbb\xi \bbb \vol(\B_r)$, we have
$$K_1(z_0) \approx \frac{\mu\big\{s\in \scal:X^{z_0}\in
\Xi^{z_0}(\B_r)\big\}}{\vol(\B_r)} \approx \int D(0,\xi,z_0)
\bbb\xi \bbb\, d\xi\,.$$ The  linear
approximation (\ref{heuristic}) leads to a similar
explanation in the case where ${\rm rank}\,V<\dim M$; we leave
this to the reader.

Before embarking on the proof of
Theorem \ref{density}, we show how the theorem 
provides a generalization of Theorem \ref{introdn} on the correlations
between zeros.  Let us first review the definition of these correlations.

\begin{defin} Let $M,V,\scal,d\mu$ be as above, and suppose that $\scal$
spans $V$. Let $M_n$ denote the punctured product (\ref{Mn}). The 
{\it $n$-point zero correlation measure\/} is the expectation $\E |Z_s|^n$,
where 
$$|Z_s|^n=\big(\underbrace{|Z_s|\times\cdots\times|Z_s|}_{n}\big)\,,$$
which is a well-defined measure on $M_n$ for almost all $s\in\scal$. We
write
$$\E |Z_s|^n=K_{n}(z^1,\dots,z^n)dz\,.$$ The generalized function
$K_{n}(z^1,\dots,z^n)$ is called the {\it $n$-point zero correlation
function\/}.
\end{defin}

\medskip We suppose $n\ge 2$ and write
$$\tilde s(z)=(s(z^1),\dots,s(z^n))\,,$$ for $ z=(z^1,\dots,z^n)\in
M_n\,,$ regarded as
a section of the vector bundle $$V_n:=\bigoplus_{p=1}^n
\pi_p^*V\longrightarrow M_n\,,$$ where $\pi_p:M_n\to M$ denotes the
projection
onto the $p$-th factor.  We then have the evaluation map
$$\ecal_n:M_n\times\scal\to V\,,\quad \ecal_n(z,s)=\tilde s(z)\,,$$ and the
jet map
$$\jcal_n:M_n \times\scal\to J^1(M_n,V_n)\,,\quad \jcal
(z,s)=J^1_z \tilde s=(J^1_{z^1}s,\dots,J^1_{z^n}s)\,.$$
We also
write
$$ x=(x^1,\dots,x^n)\in V_n\,,\ \ \xi=(\xi^1,\dots,\xi^n)\in
(T^*_M\otimes V)_{z^1}\oplus\dots\oplus(T^*_M\otimes V)_{z^n}
\subset \left(T^*_{M_n}\otimes V_n\right)_z\,,$$
$$dx=dx^1\cdots dx^n\,,\quad d\xi=d\xi^1\cdots d\xi^n\,,\quad
dz=dz^1\cdots
dz^n\,.$$ 

\begin{defin} The {\it
$n$-point
density\/} at $(z^1,\dots,z^n)\in M_n$ is the probability
measure
$$\D_n:=D_n(x,\xi,z)dxd\xi dz=\jcal_{n*} (dz\times\mu)$$ on the space
$J^1(M_n,V_n)$.  Note that this measure is supported on the sub-bundle
$$\pi_1^*(T^*_M\otimes V)\oplus\dots\oplus \pi_n^*(T^*_M\otimes V)
\subset T^*_{M_n}\otimes V_n\,.$$
The {\it ($n$-point) joint probability distribution\/} at
$(z^1,\dots,z^n)$ is the joint probability distribution
$D_n(x,\xi,z)dxd\xi=(J^1_z)_*\mu$ of the
(complex) random variables 
$$X_{jp}^z(s):=x_j(z^p)\,,\quad \Xi_{jpq}^z(s):=\xi_{jq}(z^p)\qquad
(1\le j\le k, 1\le p\le n, 1\le q
\le m)\,.$$ 
\end{defin}

If the evaluation map
$\ecal_n$ is surjective, we also write as before
\begin{equation}\label{D0n}
\D^0_n=\iota^*D(x,\xi,z)d\xi dz\,,\end{equation}
so that $$\iota_*\D^0_n = \lim_{\ep\to 0} \tilde\rho_\ep^n \D_n\,.$$
Thus, Theorem \ref{density} applied to $V_n\to M_n$ yields our general
formula for the
$n$-point correlations of zeros:

\begin{theo}\label{corr} Let $V\to M$ be a $\ccal^\infty$ vector bundle over
an oriented Riemannian manifold.  Consider the ensemble $(\scal,\mu)$, where
$\scal$ is a finite-dimensional subspace of $\ccal^\infty(M,V)$ and 
$\mu$ is given by a $\ccal^0$ rapidly decaying volume form on $\scal$.
Suppose that $\scal$ spans $V_n$, where $n$ is a positive integer.  Then
\begin{equation}\label{dn} \E|Z_s|^n =
\pi_*\left(\textstyle{\sqrt{\prod_{p=1}^n
\det(\xi^{p}\xi^{p*})}}\;\D^0\right)
\,.\end{equation}
\end{theo}
\medskip In the case where $D_n(x,\xi,z)\in\ccal^0$, (\ref{dn})
becomes
\begin{equation}\label{dn'} \E|Z_s|^n =K_n(z)dz \,,\quad
K_n(z)=
\int d\xi\,D_n(0,\xi,z)\prod_{p=1}^n
\sqrt{\det(\xi^{p}\xi^{p*})}\,.
\end{equation}

Our proof of Theorem \ref{density} uses the following {\it coarea
formula\/} of Federer:
\begin{lem} \label{Federer} {\rm \cite[3.2.12]{Fe}} Let
$f:Y\to \R^k$ be a $\ccal^\infty$ map, where $Y$ is an oriented
$m$-dimensional Riemannian manifold.  For
$\ga\in\lcal^1(Y)$, we have
$$\int_{\R^k}dx_1\cdots dx_k\int_{f\inv(x)}\ga d\vol_{m-k}=\int_Y \ga
\|df_1\wedge\cdots\wedge df_k\|d\vol_Y\,.$$
\end{lem}
\noindent  Recall that by Sard's theorem, $f\inv(x)$ is an
$(m-k)$-dimensional submanifold for almost
all $x\in\R^k$.

As a
consequence of Lemma \ref{Federer}, for $\psi\in\ccal^0(\R^k)$ we have
\begin{equation}\label{coarea} \int_{\R^k}
\,\psi(x)|f\inv(x)|\,dx_1\cdots dx_k=(\psi\circ f)
\|df_1\wedge\cdots\wedge df_k\|d\vol_Y \in\dcal'{}^m(Y)\,,\end{equation}
where $|f\inv(x)|$ denotes
$(m-k)$-dimensional volume measure on $f\inv(x)$.

\begin{rem} Federer's coarea formula, which is actually valid for
Lipschitz maps, can be regarded as in integrated form of the {\it Leray
formula}
$$|f\inv(x)|= \|df_1\wedge\cdots\wedge df_k\| \left.\frac{d\vol_Y}
{df_1\wedge\cdots\wedge df_k}\right|_{f\inv(x)}\,.$$
\end{rem}

\medskip\noindent{\it Proof of Theorem \ref{density}:\/} We restrict to a
neighborhood $U$ of an arbitrary point $z_0\in M$. Since $\scal$ spans
$V$, we can choose $U$ so that there exist sections $e_1,\dots,e_k\in\scal$
that form a local frame for $V$ over $U$. Since $\D^0$ is independent of
the connection, we can further assume that $\nabla|_U$ is the flat
connection
$\nabla s = \sum ds_j
\otimes e_j$.

For a section $s\in\scal$, we write $s(z)=\sum_{j=1}^k  s_j(z)e_j(z)$
($z\in U$) and we let $\hat s=(s_1,\dots,s_k):U\to \R^k$.  Then
$$\bbb\nabla s\bbb =\sqrt{H} \|d s_1\wedge\cdots\wedge d s_k\|\,.$$
Thus by (\ref{coarea}),
\begin{equation}\label{bycoarea} \int_{\R^k}\rho_\ep(x)|\hat s\inv(x)|dx =
(\rho_\ep\circ s)\bbb\nabla s\bbb dz\in\dcal'{}^m(U)\,,\end{equation}
where we write, as before, $dx=\sqrt{H(z)}dx_1\cdots dx_k$.

Let $\pi_U,\pi'$ denote the projections given in the commutative diagram:
$$\begin{array}{c}U\times \scal{\buildrel {\jcal}\over \to}
J^1(U,V)  {\buildrel {\iota}\over\leftarrow}  T^*_U\otimes V\\[6pt]
\pi_U\searrow\ \ \ \ \downarrow
\pi'
\ \swarrow \pi\\[6pt] U \end{array}$$
Integrating (\ref{bycoarea}) over $\scal$, we obtain
\begin{eqnarray} \int_{\R^k}\rho_\ep(x)\E|\hat s\inv(x)|dx &=& \pi_{U*}
(\rho_\ep\circ s \; \bbb\nabla s\bbb\; dz \times \mu)\nonumber \\
&=& \pi'_*\big(\rho_\ep(x)\bbb\xi\bbb\D\big)\nonumber \\
&\to & \pi'_*\big(\bbb\xi\bbb \iota_*\D^0\big) \ =\ \pi_*(\bbb\xi\bbb
\D^0)\,.\label{aa}\end{eqnarray}

To complete the proof of Theorem \ref{density}, it suffices to
show that the map
$$\Psi:\R^k
\to
\dcal'{}^m(U)\,,\quad \Psi(x)=
\E|\hat s\inv(x)|$$ is continuous; i.e., for all test functions
$\phi\in\dcal(U)$, the map $x\mapsto
\E(|\hat s\inv(x)|,\phi)$ is continuous.  Indeed, if $\Psi$ is continuous,
then
$$\left(\int_{\R^k}\rho_\ep(x)\E|\hat s\inv(x)|dx,\phi\right)=
\int \E(|\hat s\inv(x)|,\phi)\rho_\ep(x)dx \to \E(|\hat s\inv(0)|,\phi)
=\E(|Z_s|,\phi)\,,$$  and (\ref{d1}) follows from (\ref{aa}).

To verify the continuity of $\Psi$, we extend $\{e_1,\dots,e_k\}$ to a
basis $\{e_1,\dots,e_k, \dots,e_d\}$ of $\scal$, and we write
$$d\mu(s) = \psi(c_1,\dots,c_d)dc \,,\quad s=\sum_{i=1}^d c_i e_i\,.$$
We note that $$\hat s\inv(x_1,\dots,x_k)= Z_{\left[s-\sum_1^kx_je_j\right]}
\,,$$ and therefore
$$\Psi(x_1,\dots,x_k)= \int
|Z_s|\psi(c_1+x_1,\dots,c_k+x_k,c_{k+1},\dots,c_d)dc\,.$$
Write $c+x=(c_1+x_1,\dots,c_k+x_k,c_{k+1},\dots,c_d)$. We let
$\tau:I\to\R^d$ denote the projection given by $$\tau(z,\sum
c_ie_i)=(c_1,\dots,c_d)\,.$$  For a test function
$\phi\in\dcal(U)$, we have
\begin{eqnarray}(\Psi(x),\phi) &=&\int_{\R^d}(|Z_s|,\phi)\psi(c+x)dc\ =\
\int_{\R^d} \left(|\tau\inv(c)|,\phi(z)\right)\psi(c+x)dc\nonumber\\
&=& \int_I\phi(z)\psi(c+x)\|dc_1\wedge\cdots\wedge dc_d\|_I d\vol_I(z,c)
\,,\label{coareatau}\end{eqnarray} where the last equality is by the coarea
formula (\ref{coarea}) applied to $\tau$.

Suppose that $x^\nu \to x^0\in\R^k$.  In order to use (\ref{coareatau}) to
show that $(\Psi(x^\nu),\phi)\to (\Psi(x^0),\phi)$, we note that
$\|dc_1\wedge\cdots\wedge dc_d\|_I\le 1$ and hence
$$\phi(z)\psi(c+x^\nu)\|dc_1\wedge\cdots\wedge dc_d\|_I \le
\phi(z)\ga(\|c\|-R)\,,$$ where
$$\ga(r)=\sup_{\|c\|\ge r} \psi(c)\,,\qquad R=\sup_\nu\|x^\nu\|\,.$$
We let $I(r)=\left\{(z,\sum c_ie_i)\in I:\|c\|=r\right\}$ denote the sphere
bundle of radius $r$ in the vector bundle $I\to M$.  We then have
$$\int_I\phi(z)\ga(\|c\|-R)d\vol_I(z,c)=\int_0^{+\infty} dr\;
\ga(r-R)\int_{I(r)}\phi(z)d\vol_{I(r)}=  C \int_0^{+\infty}
dr\;\ga(r-R) r^{d-1}
\,.$$  Since by hypothesis $\ga(r)=o(r^{-d-1})$, we conclude that the
integral is finite and thus the Lebesgue
dominated convergence theorem implies that $(\Psi(x^\nu),\phi)\to
(\Psi(x^0),\phi)$.  \qed

\subsection {Zero correlations on complex manifolds}
We now describe the jet density $\D$ in the case where $(V,h)$ is a
complex
hermitian vector bundle.  In this case, we choose a complex local frame
$\{e_1,\dots,e_k\}$ and we let $H_\C=\det(h_{jj'}),\ h_{jj'}=h(e_j,e_{j'})$.
We write $$x=\sum_j x_je_j\,,\quad\xi=\sum_{j,q} \xi_{jq} du_q\otimes
e_j=\sum_j
\xi_j\otimes e_j\,,$$ where $\xi_{jq},\ x_j$ are complex.  We then have
$$\D=D(x,\xi,z)dx d\xi dz\,,$$ where this time $$dx=H_\C(z)
\prod_{j} d\Re x_j d\Im x_j\,, \quad d\xi=
G(z)^{-k/2}H_\C(z)^{m} \prod_{j,q} d\Re \xi_{jq} d\Im
\xi_{jq}\,.$$ We also have
$$\bbb\xi\bbb = H_\C\|\xi_1\wedge\cdots\wedge\xi_k \wedge
\bar\xi_1\wedge\cdots\wedge\bar\xi_k\|\,.$$

We can now specialize Theorems \ref{density}--\ref{corr} to the case where
$V$ is a holomorphic line bundle over
a complex manifold $M$ and the
sections in $\scal$ are holomorphic.  If we now let $\{z_q\}$ denote complex
local coordinates, we can write
$$\xi=\xi'+\xi''= \sum_{j,q} (\xi_{jq}' dz_q + \xi_{jq}'' d\bar z_q)\otimes e_j
\,.$$ Since
$\dbar s=0$ for all $s\in\scal$, the support of the measure $\D$ is
contained
in $V\oplus (T^{*h}_M\otimes V)$, i.e., those $(x,\xi)$ with $\xi_{jq}''\equiv
0$ (using a holomorphic frame  $\{e_j\}$ and a connection $\nabla$ of type
(1,0)).  Hence on the support of $\D$, we have $\xi_j\in T^{*h}_M$, and
hence
\begin{equation}\label{holo} \bbb\xi\bbb =
H\|\xi_1\wedge\cdots\wedge\xi_k\|^2
=\det(\xi\xi^*)_\C\,,\end{equation} where $(\xi\xi^*)_\C\in \mbox{\rm
Hom}_\C(V_z,V_z)$ denotes the complex endomorphism.  Hence as a special case
of Theorem \ref{corr}, we obtain:

\begin{theo}\label{corrcx} Let $V\to M$ be a holomorphic line bundle over a
complex manifold $M$ and let $\scal$ be a finite dimensional complex
subspace of $H^0(M,V)$.  We give $\scal$ a semi-positive rapidly decaying
volume form $\mu$.  If $\scal$ spans $V_n$, then
\begin{equation}\label{dncx} \E|Z_s|^n =
\pi_*\left({\textstyle \prod_{p=1}^n}\det(\xi\xi^*)_\C\,\D^0\right)
\,.\end{equation}
\end{theo}
\noindent In the case where the image of $\jcal_n$ contains all the
holomorphic 1-jets, we can write $\D_n=D_n(x,\xi',z)dxd\xi' dz$,
$D_n(x,\xi,z)\in\ccal^0$. Then (\ref{dncx}) yields the following
result from
\cite[Th.~2.1]{BSZ2}:
\begin{equation}\label{dn'cx} \E|Z_s|^n =K_n(z)dz \,,\quad
K_n(z)=
\int d\xi\,D_n(0,\xi,z)\prod_{p=1}^n
{\det(\xi^{p}\xi^{p*})_\C}\,.
\end{equation}

\section {Universality of the scaling limit of the
correlations}\label{s-universality}

We return
to
our complex Hermitian line bundle $(L,h)$ on a compact almost complex
$2m$-dimensional symplectic manifold
$M$
with symplectic form $\omega=\frac{i}{2}\Theta_L$, where $\Theta_L$ is the
curvature of $L$ with respect to a connection $\nabla$. Theorem
\ref{introdn} follows from Theorem \ref{corr} applied to the vector bundle
$$V=\underbrace{L^N\oplus\cdots\oplus L^N}_k$$
and the (finite-dimensional) space of sections
$$\scal
=H^0_J(M,L^N)^k \subset \ccal^\infty(M,V)\,.$$

\subsection{Gaussian measures}\label{s-gaussians}

Recalling (\ref{inner}), we consider the
Hermitian inner product  on\break $H^0_J(M,L^N)$:
\begin{equation*}\label{inner2}\langle s_1, s_2 \rangle = \int_M h^N(s_1,
s_2)\frac{1}{m!}\om^m \quad\quad (s_1, s_2 \in
H^0_J(M,L^N)\,)\,.\end{equation*}
We give
$\scal$ the Gaussian probability measure
$\mu_N=\nu_N\times\cdots\times\nu_N$, where
$\nu_N$ is the `normalized' complex Gaussian measure on $H^0_J(M,L^N)$:
\begin{equation}\label{cxgaussian}\nu_N(s)=
\left(\frac{d_N}{\pi}\right)^{d_N}e^
{-d_N|c|^2}dc\,,\qquad s=\sum_{j=1}^{d_N}c_jS_j^N\,.\end{equation} Here
$\{S_j^N\}$ is an orthonormal basis for $H^0_J(M,L^N)$ (with respect to the
Hermitian inner product (\ref{inner})) and $dc$ is
$2d_N$-dimensional Lebesgue measure.  The normalization is chosen so
that $\E\langle s,s\rangle=1$.  This Gaussian  is characterized by the
property that the
$2d_N$ real variables
$\Re c_j, \Im c_j$ ($j=1,\dots,d_N$) are independent identically distributed
(i.i.d.) random variables with mean 0 and variance $\frac{1}{2d_N}$;
i.e.,
$$\E c_j = 0,\quad \E c_j c_k = 0,\quad  \E c_j \bar c_k =
\frac{1}{d_N}\de_{jk}\,.$$
Picking a random element of $\scal$ means picking $k$
sections of
$H^0_J(M,L^N)$ independently and at random.

\begin{rem}
Since we are interested in the zero sets $Z_s$, which do not depend on
constant factors, we could just as well suppose our sections lie in the unit
sphere $SH^0_J(M,L^N)$ with respect to the Hermitian inner product
(\ref{inner}), and pick random sections with respect to the spherical 
measure.
This gives the same expectations for $|Z_s|^n$ as the Gaussian measure on
$H^0_J(M,L^N)$. \end{rem}

We now review the concept of `generalized Gaussian measures' from
\cite{SZ2}, which is one of the ingredients in obtaining the (universal)
scaling limit of the joint probability distribution, which in turn yields
the universality of the scaling limit of the correlation of zeros on
symplectic manifolds.  (For further details and related results, see
\cite[\S 5.1]{SZ2}.) To begin, a (non-singular) Gaussian measure $\ga_\De$ on
$\R^p$ given by (\ref{gaussian})
has second moments
\begin{equation}\label{moments}\langle x_jx_k
\rangle_{\ga_\De}=\De_{jk}\,.\end{equation} The measure $\ga_\De$ is
characterized by its Fourier transform
\begin{equation}\label{muhat}\wh{\ga_\De}(t_1,\dots,t_p) = e^{-\half\sum
\De_{jk}t_jt_k}\,.\end{equation}

The push-forward of a Gaussian measure by a surjective linear map is also
Gaussian. Since we need to push forward Gaussian
measures (on the spaces
$H^0_J(M,L^N)$) by linear maps that are sometimes not surjective, we shall
consider the case where
$\De$ is positive semi-definite.  In this
case, we can still use (\ref{muhat}) to define a measure $\ga_\De$, which we
call a {\it generalized Gaussian\/}.  If $\De$ has null eigenvalues, then
$\ga_\De$ is a Gaussian measure on the subspace $\Lambda_+\subset\R^p$
spanned by the positive eigenvectors.  
If
$\ga$ is a generalized Gaussian on $\R^p$ and
$L:\R^p\to\R^q$ is a (not necessarily surjective) linear map, then $L_*\ga$
is a generalized Gaussian on $\R^q$.  By studying the Fourier transform, it 
is
easy to see that the  map $\De\mapsto\ga_\De$ is a continuous
map from the positive semi-definite matrices to the space of positive
measures on
$\R^p$ (with the weak topology).

\subsection{Densities and the Szeg\"o kernel}\label{D-meet-S}

We now consider  the $n$-point joint probability distribution
of a (Gaussian) random almost holomorphic
section
$s\in H^0_J(M,L^N)$ having prescribed values $s(z^p) = x^p$ and
prescribed derivatives
$\nabla s(z^p) =\xi^p$ (for $1\le p\le n$). We denote this
density by $\wt D_{n}^N(x,\xi,z)dxd\xi$ as in \cite{SZ2}, where
$z=(z^1,\dots,z^n)$.  Having equipped
$H^0_J(M,  L^N)$ with the Gaussian measure $\nu_N$, and recalling that the 
joint probability distribution
$$\wt\D_{z}^N:=
\wt D_{n}^N(x,\xi,z)dxd\xi=(J_z^1)_*\nu_N\,,
$$ is the push-forward of $\nu_N$ by a linear map,
we conclude that the joint probability distribution is  a generalized 
Gaussian measure  on the complex vector space of 1-jets of sections:
$$\wt\D_{z}^N=\ga_{\De^N(z)}\,.$$ 

To be more precise,  we consider the
$n(2m+1)$  complex-valued random
variables $X_p,\ \Xi_{pq}$ ($1\le p\le n,\ 1\le q\le 2m$)
on $\hcal^2_N(X)\equiv H^0_J(M,L^N)$ given by 
\begin{equation}\label{dms1def} X_p(s) = s(z^p,0)\,,\quad \Xi_{pq}(s)=
(\nabla_q) s(z^p,0)\,,
\end{equation} where \begin{equation}
\label{dms2def} \nabla_q=\frac{1}{\sqrtn} \frac{\d^h }{\d
z_q}\,,\quad
\nabla_{m+q}=\frac{1}{\sqrtn} \frac{\d^h }{\d\bar z_q}\qquad
(1\le q\le m)\,,
\end{equation}
for $s\in \hcal^2_N(X)$.  Here, $\d^h/\d z_q$ denotes the horizontal
lift to $X$ of the tangent vector $\d/\d z_q$ on $M$.  The  covariance
matrix $\De^N(z)$ is given by the Szeg\"o kernel and its covariant
derivatives, as follows:

\begin{eqnarray}
\Delta^N(z)&=&\left(
\begin{array}{cc}
A^N & B^N \\
B^{N*} & C^N
\end{array}\right)\,,\nonumber \\
\big( A^N\big)^{p}_{p'} &=&
\E\big ( X_p \bar
X_{p'}\big)=\frac{1}{d_N}\Pi_N(z^p,0;z^{p'},0)\,,\nonumber\\
\big( B^N\big)^{p}_{p'q'}&=&
\E\big( X_p \overline \Xi_{p'q'}\big)= \frac{1}{d_N}
\overline{\nabla}^2_{q'}\Pi_N(z^p,0;z^{p'},0)\,,\nonumber\\
\big( C^N\big)^{pq}_{p'q'}&=&
\E\big(  \Xi_{pq}\overline \Xi_{p'q'}\big)=
 \frac{1}{d_N}
\nabla^1_q\overline{\nabla}^2_{q'}\Pi_N(z^p,0;z^{p'},0)\,,\nonumber\\
&& p,p'=1,\dots,n, \quad q,q'=1,\dots,2m\,.\nonumber
\end{eqnarray}
Here, $\nabla^1_q$, respectively $\nabla^2_q$, denotes
the differential operator on $X\times X$ given by applying
$\nabla_q$ to the first, respectively second, factor.
(We note that $A^N,\ B^N,\ C^N$ are $n\times n,\ n\times 2mn,\
2mn\times 2mn$ matrices, respectively; $p,q$ index the rows, and
$p',q'$ index the columns.)
In \cite{BSZ2} we proved
that the joint  probability density has a universal scaling limit,
and in
\cite{SZ}  this  result was extended to the  symplectic case:

\begin{theo}\label{usljpd-sphere} {\rm (\cite{SZ2}, Theorem 5.4)} Let
$L$ be a pre-quantum line bundle over  a
$2m$-\break dimensional compact integral symplectic manifold $(M,\om)$.
Choose Heisenberg coordinates $\{z_j\}$ about a point
$P_0\in M$. Then
$$\wt \D^N_{(z^1/\sqrtn,\dots, z^n/\sqrtn)} \longrightarrow \D^\infty
_{(z^1,\dots,n^n)}= \ga_{\De^\infty(z)}$$   where
$\D^\infty_{(z^1,\dots,z^n)}$ is a universal Gaussian measure supported on
the holomorphic 1-jets, and $\De^N(z/\sqrtn)\to \De^\infty(z)$.
\end{theo}

Theorem \ref{usls} then follows immediately from  Theorems  
\ref{introdn} and \ref
{usljpd-sphere}. In fact, we have the error estimate
$$\left(\frac{1}{ N^{nk}} K_{nk}^N\left({\textstyle \frac{z^1}{\sqrtn},
\dots,
\frac{z^n}{\sqrtn}}\right), \phi\right) =
\left(K_{nkm}^\infty(z^1,\dots,z^n),\phi\right) +
O\left(\frac{1}{\sqrtn}\right)\,,$$ for all $\phi\in\dcal^{mn}((\C^m)_n)$.

A technically interesting novelty in the proof of Theorem
\ref{usljpd-sphere} is the role of the $\bar{\partial}$ operator. In the
holomorphic case,
$\wt\D^N_{(z^1,\dots,z^n)}$ is supported
on the subspace of sections satisfying $\bar{\partial} s = 0$.  In the
almost
complex case,  sections do not satisfy this equation, so
$\wt\D^N_{(z^1,\dots,z^n)}$ is a measure on a  higher-dimensional space of
jets. However, Theorem \ref{usljpd-sphere} says that the mass in the
`$\bar{\partial}$-directions' shrinks to zero as $N \to \infty$.

An alternate statement of Theorem \ref{usljpd-sphere} involves equipping
the unit spheres $H^0(M, L^N)$ with Haar probability measure, and letting 
$\D^N_{(z^1,\dots,z^n)}$
be the corresponding joint probability distribution on $SH^0_J(M, L^N)$.
In \cite[Theorem~0.2]{SZ}, it was shown that these 
non-Gaussian measures
$\D^N$ also have the same scaling limit $\D^\infty$. 

The matrix $\De^\infty$ is given in terms of the \szego kernel for the
Heisenberg group:
\begin{equation}\label{delta}
\De^\infty(z)= \frac{m!}{c_1(L)^m}\left(
\begin{array}{cc}
A^\infty(z) & B^\infty(z) \\
B^{\infty}(z)^* & C^\infty(z)
\end{array}\right)\,,\end{equation} where
\begin{eqnarray*}
 A^\infty(z)^p_{p'} &=& \Pi_1^\H(z^p,0;z^{p'},0)\,,\\
B^\infty(z)^{p}_{p'q'} &=&\left\{\begin{array}{ll}
(z^p_{q'}-z^{p'}_{q'}) \Pi_1^\H(z^p,0;z^{p'},0) \  &\mbox{for}\quad  1\le
q\le m\\ 0 & \mbox{for}\quad  m+1\le q\le 2m\end{array}\right.\ ,\\
C^\infty(z)^{pq}_{p'q'} &=&\left\{\begin{array}{ll}
\left(\delta_{qq'}+(\bar z^{p'}_q
-\bar z^p_q) (z^p_{q'}-z^{p'}_{q'})\right)\Pi_1^\H(z^p,0;z^{p'},0) \ 
&\mbox{for}\quad  1\le q,q'\le m\\ 0 & \mbox{for}\quad  \max(q,q')\ge
m+1\end{array}\right.\ .
\end{eqnarray*}
For details, see \cite{SZ2}. 

Equation (\ref{delta}) says that the
variances in the anti-holomorphic directions vanish.  If we remove the rows
and columns of the matrices corresponding to $m+1\le q \le 2m$, then we get
the covariance matrix 
\begin{equation}\label{delta-h}
\De_h^\infty(z)= \frac{m!}{c_1(L)^m}\left(
\begin{array}{cc}
A^\infty(z) & B_h^\infty(z) \\
B_h^{\infty}(z)^* & C_h^\infty(z)
\end{array}\right)\end{equation}
for the joint probability distribution in the holomorphic case. (Here
$A^\infty,\ B_h^\infty,\ C_h^\infty$ are $n\times n,\ n\times mn,\
mn\times mn$ matrices, respectively.)  In \cite{BSZ2}, we used (\ref{dn'cx})
and (\ref{delta-h}) to obtain  formulas for the scaling limit zero
correlations
$K^\infty_{nkm}$. We briefly summarize here how it was done: 
Let us write $$\D^\infty_{(z^1,\dots,z^n)}=\D^\infty_z =
D^\infty(x,\xi,z)dxd\xi\,.$$  The function $D^\infty(0,\xi,z)$ is Gaussian in
$\xi$, but is not normalized as a probability density.  It is given by
\begin{equation}\label{D0infty} D^\infty(0,\xi,z)d\xi = \frac{1}{\pi^n\det
A^\infty(z)}\ga_{\La^\infty(z)}
\,,\end{equation} where 
\begin{equation}\label{La}
\La^\infty(z) = C_h^\infty(z) -B_h^\infty(z)^* A^\infty(z)\inv
B_h^\infty(z)\,.
\end{equation}

We first consider the  $k=1$ case of the limit correlation function
 for the zero divisor (complex hypersurface) of one random
section. By (\ref{dn'cx}), (\ref{D0infty}), and the identity $\det
\De^\infty_h=\det \La^\infty \det A^\infty$, we obtain
\begin{equation}\label{Kn1m} K^\infty_{n1m}(z^1,\dots,z^n)=
\frac{1}{\pi^n\det A^\infty(z)} \int_{\C^{mn}} \prod_{p=1}^n \left(
\sum_{q=1}^m
|\xi_q^p|^2\right)d\ga_{\La^\infty(z)}(\xi) \,.\end{equation}  The integral
in (\ref{Kn1m}) is a sum of
$(2n)^{\rm th}$ moments of the Gaussian measure $\ga_{\La^\infty(z)}$, and
can be evaluated using the Wick formula.  Indeed, in the pair correlation
case $n=2$, (\ref{Kn1m}) yields the explicit formula (\ref{paircor}).

For the case of random $k$-tuples $s=(s^1,\dots,s^k)\in \scal=H^0_J(M,L^N)^k$
(where the zero sets are of codimension $k$), the 1-jets
$J^1_z\tilde s^1,\dots, J^1_z\tilde s^k$ are i.i.d.\ random vectors, and we
have
\begin{equation}\label{Knkm} K^\infty_{nkm}(z^1,\dots,z^n)=
\frac{1}{[\pi^{n}\det A^\infty(z)]^k}\int_{\C^{kmn}} \prod_{p=1}^n
\det_{1\le j,j'\le k}\left(
\sum_{q=1}^m \xi_{jq}^p \bar \xi_{j'q}^p
\right)d\ga_{I_k\otimes\La^\infty(z)}(\xi)\,,\end{equation}
where $I_k$ denotes the $k\times k$ identity matrix; i.e., 
$$\big( I_k\otimes\La^\infty(z) \big)^{jpq}_{j'p'q'} = \de^j_{j'}
\La^\infty(z)^{pq}_{p'q'}\,.$$  
For further details and explicit formulas, see
\cite{BSZ2}  for the case $k=n=2$, and see \cite{BSZ3} for the point
pair correlation case $n=2,\ k=m$. Indeed, we show  in
\cite{BSZ3} that for small values of $r:=|z^1-z^2|$, we have
$$ \widetilde K_{2mm}(z^1,z^2)= \frac{m+1}{4}
r^{4-2m} + O(r^{8-2m})\,,\qquad m=1,2,3,\dots\,.$$

\subsection{Decay of correlations}\label{decay} 
Let us define the {\it normalized $n$-point scaling limit zero
  correlation function}
\begin{equation}\label{nslzc} \wt K^\infty_{nkm}(z)=
(K^\infty_{1km})^{-n}K^\infty_{nkm}(z)=\left(\frac{\pi^k(m-k)!}{m!}\right)^n
K^\infty_{nkm}(z)\,.\end{equation}
In \cite{BSZ2}, we showed that the limit
correlations are ``short range" in the following sense:

\begin{theo}\label{shortrange} {\rm (\cite{BSZ2}, Theorem 4.1)} The
correlation functions satisfy the estimate
$$\wt
K^\infty_{nkm}(z^1,\dots,z^n) = 1 +O(r^4 e^{-r^2}) \quad {\rm as}\ r\to
\infty\,, \quad r=\min_{p\ne p'}|z^p-z^{p'}|\,.$$ \end{theo}

We review here the proof of this estimate.  Writing
$$A=\pi^mA^\infty\,,\quad
B=\pi^mB^\infty_h\,,\quad
C=\pi^mC^\infty_h\,,\quad \La=\pi^m\La^\infty\,,$$ we have:
\begin{eqnarray}\label{ABC} A^p_{p'}&=& 
e^{ i \Im (z^p \cdot \bar z^{p'})} e^{- 
\half |z^p - z^{p'}|^2}\,,\nonumber \\
B^{p}_{p'q'} &=&  (z^{p}_{q'}-z^{p'}_{q'}) A^p_{p'}\,,\\
C^{pq}_{p'q'} &=&  \big[\de_{qq'}+(\bar z^{p'}_q-\bar z^{p}_{q})
( z^{p}_{q'}-z^{p'}_{q'})\big] A^p_{p'}
\,.\nonumber \end{eqnarray}
This implies that
\begin{equation}\label{ABC2} 
\begin{aligned}
A&=\,I+O(
e^{-r^2/2})\,,\qquad  A^{p}_{p}=1\,,\\
B&=\,O(re^{-r^2/2})\,,\\
C&=\,I+O(r^2
e^{-r^2/2})\quad \mbox{as}\ r\to\infty\,,\qquad   C^{pq}_{pq}=1
\,.
\end{aligned}
\end{equation}
Recalling (\ref{La}), we have
\begin{equation}\label{Lambda}\La=I+O(r^2 e^{-r^2/2})\,,\qquad
\La^{pq}_{pq}=1+O(r^2 e^{-r^2})\,,
\qquad \mbox{as}\ r\to\infty\,.\end{equation} We now use the
Wick formula to evaluate the integral in (\ref{Knkm}). (Formula (\ref{Knkm})
is homogeneous of order 0 in the matrix entries, so is not affected when
$A^\infty,\,\La^\infty$ are multiplied by $\pi^m$.)  Note that the Wick
formula involves terms that are  products of diagonal elements of
$\La$, and products that contain at least two off-diagonal elements of
$\La$.  The former terms are of the form $1+O(r^2e^{-r^2})$, and the latter
are
$O(r^4e^{-r^2})$. Similarly, $\det A=1+O(e^{-r^2})$, and the
estimate follows. \qed

\bigskip The theorem can be extended to
estimates of the connected  correlation functions (called also truncated
correlation functions, cluster functions, or cumulants), as follows. The
$n$-point connected correlation function is defined as
(see, e.g., \cite[p.~286]{GJ}) 
\begin{equation}\label{cc1}\wt
T^{\infty}_{nkm}(z^1,\dots,z^n)=\sum_G(-1)^{l+1}(l-1)!
\prod_{j=1}^l \wt K^{\infty}_{n_jkm}(z^{p_{j1}},\dots,z^{p_{jn_j}}),
\end{equation}
where the sum is taken over all partitions $G=(G_1,\dots,G_l)$ of the
set $(1,\dots,n)$ and $G_j=(p_{j1},\dots,p_{jn_j})$. In particular, recalling that
$\wt K^\infty_{1km}\equiv 1$,
\begin{eqnarray*}\label{cc2}
\wt T^{\infty}_{1km}(z^1)&=&\wt K^{\infty}_{1km}(z^1)=1\,,\\
\wt T^{\infty}_{2km}(z^1,z^2)&=&\wt K^{\infty}_{2km}(z^1,z^2)
-\wt K^{\infty}_{1km}(z^1)\wt K^{\infty}_{1km}(z^2)
\ =\ \wt K^{\infty}_{2km}(z^1,z^2)-1\,,\\
\wt T^{\infty}_{3km}(z^1,z^2,z^3)&=&\wt K^{\infty}_{3km}(z^1,z^2,z^3)
-\wt K^{\infty}_{2km}(z^1,z^2)\wt K^{\infty}_{1km}(z^3)
-\wt K^{\infty}_{2km}(z^1,z^3)\wt K^{\infty}_{1km}(z^2)\\
&&\ -\ \wt K^{\infty}_{2km}(z^2,z^3)\wt K^{\infty}_{1km}(z^1)
+2 \wt K^{\infty}_{1km}(z^1)\wt K^{\infty}_{1km}(z^2)
\wt K^{\infty}_{1km}(z^3)\\
&=&\wt K^{\infty}_{3km}(z^1,z^2,z^3)
-\wt K^{\infty}_{2km}(z^1,z^2) -\wt K^{\infty}_{2km}(z^1,z^3) 
-\wt K^{\infty}_{2km}(z^2,z^3)+2\,, \end{eqnarray*}
and so on. The inverse of (\ref{cc1}) is
\begin{equation}\label{cc3}\wt
K^{\infty}_{nkm}(z^1,\dots,z^n)=\sum_G
\prod_{j=1}^l \wt T^{\infty}_{n_jkm}(z^{p_{j1}},\dots,z^{p_{jn_j}})
\end{equation}
(M\"obius theorem).
The advantage of the connected correlation functions is that they
go to zero if at least one of the distances $|z^i-z^j|$ goes to infinity (see
Corollary \ref{ccc} below).
In our case the connected correlation functions can be estimated as
follows. Define
\begin{equation}\label{cc4}
d(z^1,\dots,z^n)=\max_{\gcal}\prod_{l\in L}
|z^{i(l)}-z^{f(l)}|^2e^{-|z^{i(l)}-z^{f(l)}|^2/2}.
\end{equation}
where the maximum is taken over all oriented connected graphs
$\gcal=(V,L)$ ``with zero boundary" such that $V=(z^1,\dots,z^n)$. Here $V$
denotes the set of vertices of $\gcal$, $L$ the set of edges, and $i(l)$
and $f(l)$ stand for the initial and final vertices of the edge $l$,
respectively. The graph
$\gcal$ is said to be have zero boundary if $\sum \{l:l\in L\}$ is a
1-cycle; i.e., for each vertex $z^p\in V$, the
number of edges beginning at  $z^p$ equals the number ending at $z^p$.
(There must be at least one edge beginning at
each vertex, since
$\gcal$ is assumed to be connected.  Graphs may have any number of edges
connecting the same two vertices.) Observe that
the maximum in (\ref{cc4}) is achieved at some graph $\gcal$, because
$te^{-t/2}\le 2/e<1$ and therefore the product in (\ref{cc4}) is at most
$(2/e)^{|L|}$.

\begin{theo}\label{cc} The connected correlation functions satisfy the estimate
$$\wt T^\infty_{nkm}(z^1,\dots,z^n) = O(d(z^1,\dots,z^n)) \,,$$ provided that
$\min_{p\ne q}|z^p-z^q|\ge c>0$. \end{theo}

To prove the theorem, let us introduce the $n$-point functions
\begin{equation}\label{Khat}
\begin{aligned}
\wh K_n(z^1,\dots,z^n)
&=\det A_{nkm}(z^1,\dots,z^n)\;\wt K^{\infty}_{nkm}(z^1,\dots,z^n)\\
&=[(m-k)!/m!]^n\int
\prod_{p=1}^n
\det_{1\le j,j'\le k}\left(
\sum_{q=1}^m \xi_{jq}^p \bar \xi_{j'q}^p
\right) d\ga_{I_k\otimes\La(z)}(\xi)
\end{aligned}
\end{equation}
where $A_{nkm}=I_k\otimes A$, an $nk\times nk$ matrix. 
(Note that
$\det  A_{nkm}=(\det A)^k$.  It was shown in \cite[Lemma~3.3]{BSZ2}
that
$\det A>0$ at distinct points $z^p$.) We also consider the corresponding
``connected functions"
\begin{equation}\label{cchat}\wh
T_n(z^1,\dots,z^n)=\sum_G(-1)^{l+1}(l-1)!
\prod_{j=1}^l \wh K_{n_j}(z^{p_{j1}},\dots,z^{p_{jn_j}})\,,
\end{equation} and we note that the M\"obius inversion formula applies to $\wh
K_n,\
\wh T_n$. 

Observe that we can rewrite $\wh K_n(z^1,\dots,z^n)$  as a sum over Feynman
diagrams. Namely, each term in the Wick sum for the integral in
(\ref{Khat}) corresponds to a graph $\fcal=(V,L)$ (Feynman diagram) such
that $V=(z^1,\dots,z^n)$ and the edges $l\in L$ connect the paired
variables $\xi^{i(l)}_{jq},\; \bar\xi^{f(l)}_{jq'}$ in the given Wick term.
We have that
\begin{equation}\label{cc5}
\wh K_n(z^1,\dots,z^n)
=[(m-k)!/m!]^n
\sum_{\fcal} W_{\fcal}(z^1,\dots,z^n)\,,
\end{equation}
where the function
$W_{\fcal}(z^1,\dots,z^n)$ is the sum over all terms in the Wick sum
corresponding to the
Feynman diagram $\fcal$.  (In other words, to get
$W_{\fcal}(z^1,\dots,z^n)$ we fix the indices $p,p'$ of the 
pairings $(\xi^p_{jq},\bar\xi^{p'}_{jq'})$
prescribed by $\fcal$ and sum up in the Wick formula over all indices $j,q$ at
every $z^p$.)  Note that each graph $\fcal$ in the sum
(\ref{cc5}), having arisen from a term in the Wick sum, has zero boundary.

A remarkable property of the ``connected functions" is
that they are represented by the sum over connected Feynman diagrams (see,
e.g., \cite{GJ}):
\begin{equation}\label{cc6}\wh T_n(z^1,\dots,z^n)= 
[(m-k)!/m!]^n{\sum_{\fcal}}^{\rm conn} 
W_{\fcal}(z^1,\dots,z^n)\,.\end{equation}
We conclude from  (\ref{Lambda}) that for all connected Feynman diagrams
$\fcal$,
\begin{equation}\label{cc7}
W_{\fcal}(z^1,\dots,z^n)=O(d(z^1,\dots,z^n))\,,\quad \mbox{provided that}\
\min_{p\ne q}|z^p-z^q|\ge c>0\,.
\end{equation}
Summing over
$\fcal$, we obtain the following estimate:
\begin{lem}\label{cchat2}
$\displaystyle\quad\wh T_n(z^1,\dots,z^n) = O(d(z^1,\dots,z^n))\,,
\quad \mbox{provided that}\ \
\min_{p\ne q}|z^p-z^q|\ge c>0\,.$
\end{lem}

It remains to relate $\wt T^\infty_{nkm}(z^1,\dots,z^n)$ to
$\wh T_n(z^1,\dots,z^n)$. To do this, we introduce the functions
\begin{equation}\label{ccA}
Q_n(z^1,\dots,z^n)=\sum_G(-1)^{l+1}(l-1)!
\prod_{j=1}^l  \det\, A_{n_jkm}(z^{p_{j1}},\dots,z^{p_{jn_j}}),
\end{equation}
which are the connected  functions for $\det
A_{nkm}(z^1,\dots,z^n)$, and 
\begin{equation}\label{cc1/A}
R_n(z^1,\dots,z^n)=\sum_G(-1)^{l+1}(l-1)!
\prod_{j=1}^l \frac{1}{ \det\,
  A_{n_jkm}(z^{p_{j1}},\dots,z^{p_{jn_j}})},
\end{equation}
which are the connected functions for
$\displaystyle\frac{1}{\det A_{nkm}(z^1,\dots,z^n)}$. Recall the M\"obius
inversion formula
\begin{equation} \frac{1}{\det A_{nkm}(z^1,\dots,z^n)}= \sum_G
\prod_{j=1}^l R_{n_j}(z^{p_{j1}},\dots,z^{p_{jn_j}})
\label{Rmoebius}\;.\end{equation}

We have the following relation between $\wt
T^\infty_{nkm}(z^1,\dots,z^n)$ and $\wh T_n(z^1,\dots,z^n)$. 

\begin{lem}\label{lem_t-h1} 
\begin{equation}\label{theqn}
\wt T^\infty_{nkm}(z^1,\dots,z^n)=
{\sum_{G,H}}^{\rm conn}\,
\prod_{j=1}^l  \wh T_{n_j}(z^{p_{j1}},\dots,z^{p_{jn_j}})
\prod_{j=1}^{l'}  R_{m_j}(z^{p'_{j1}},\dots,z^{p'_{jm_j}})
\end{equation}
where the sum is taken over all pairs $\{G=(G_1,\dots,G_l),\;
H=(H_1,\dots,H_{l'})\}$  of partitions of the
set $(1,\dots,n)$ which are ``mutually connected'' in the sense that
there is no proper subset $S$ of the set $(1,\dots,n)$ such that $S$
is a union of some subsets $G_j$ and  is also a union of some
subsets $H_j$. In {\rm (\ref{theqn})}, $G_j=(p_{j1},\dots,p_{jn_j})$ and
$H_j=(p'_{j1},\dots,p'_{jm_j})$. 
\end{lem}

\begin{proof} The proof is by induction on $n$. From (\ref{cc3}),
\begin{equation}\label{t-h1}
\wt T^{\infty}_{nkm}(z^1,\dots,z^n)=\wt
K^{\infty}_{nkm}(z^1,\dots,z^n)-{\sum_F}' 
\prod_{j=1}^l \wt T^{\infty}_{n_jkm}(z^{p_{j1}},\dots,z^{p_{jn_j}})
\end{equation}
where the summation goes over all partitions $F=(F_1,\dots,F_l)$ with
at least two 
elements in the partition  (i.e., $l\ge 2$). From 
(\ref{cchat}) and (\ref{Rmoebius}), we have
\begin{eqnarray}\label{t-h2}\wt
K^{\infty}_{nkm}(z^1,\dots,z^n)&=&
\wh K_n(z^1,\dots,z^n)\frac{1}{\det A_{nkm}(z^1,\dots,z^n)}\nonumber\\ &=&
\sum_{G,H}\,
\prod_{j=1}^l  \wh T_{n_j}(z^{p_{j1}},\dots,z^{p_{jn_j}})
\prod_{j=1}^{l'}  R_{m_j}(z^{p'_{j1}},\dots,z^{p'_{jm_j}})\,,
\end{eqnarray}
where the sum is taken over all pairs $\{G=(G_1,\dots,G_l),\;
H=(H_1,\dots,H_{l'})\}$  of partitions of the
set $(1,\dots,n)$. If we use the inductive assumption that
(\ref{theqn}) holds when the number of points is less than $n$ and
apply it to $\wt T^\infty_{n_jkm}(z^{p_{j1}},\dots,z^{p_{jn_j}})$ in
(\ref{t-h1}), then we obtain that
\begin{equation}\label{t-h3}
{\sum_F}' 
\prod_{j=1}^l \wt T^{\infty}_{n_jkm}(z^{p_{j1}},\dots,z^{p_{jn_j}})
={\sum_{G,H}}^{\rm disconn}\,
\prod_{j=1}^l  \wh T_{n_j}(z^{p_{j1}},\dots,z^{p_{jn_j}})
\prod_{j=1}^{l'}  R_{m_j}(z^{p'_{j1}},\dots,z^{p'_{jm_j}})\,,
\end{equation}
where the sum on the right 
is taken over all partitions $G=(G_1,\dots,G_l)$ of the
set $(1,\dots,n)$ and all partitions $H=(H_1,\dots,H_p)$ of the
set $(1,\dots,n)$ which are ``mutually disconnected'' in the sense that
there is a proper subset $S$ of the set $(1,\dots,n)$ that
is simultaneously a union of some subsets $G_j$ and  a union of
some subsets $H_j$. When we substitute (\ref{t-h2}) and (\ref{t-h3}) into
(\ref{t-h1}) and take the difference on the right of
(\ref{t-h1}), disconnected pairs $\{G,H\}$ will be cancelled out and
we will be left with mutually connected $\{G,H\}$.  This proves the lemma.
\end{proof}

\begin{lem}\label{lem_t-h2}
 The functions $ Q_n(z^1,\dots,z^n)$ satisfy the estimate
\begin{equation}\label{t-h4}
 Q_n(z^1,\dots,z^n)=O(d(z^1,\dots,z^n))\,,
\end{equation} provided that
$\min_{p\ne q}|z^p-z^q|\ge c>0$.
\end{lem}

\begin{proof} By the determinant formula,
\begin{equation}\label{t-h5}
\det A_{nkm}(z^1,\dots,z^n)= (\det A)^k=\sum_\pi
(-1)^{\sigma(\pi)}\prod_{j=1}^k\prod_{p=1}^n A_p^{\pi_j(p)}\,, 
\end{equation}
where the sum goes over all $k$-tuples $\pi=
(\pi_1,\dots,\pi_k)$ of permutations  of
$(1,\dots,n)$. 
We claim that 
\begin{equation}\label{t-h6}
 Q_n(z^1,\dots,z^n)={\sum_\pi}^{\rm conn}
 (-1)^{\sigma(\pi)}\prod_{j=1}^k\prod_{p=1}^n A_p^{\pi_j(p)}\,,
\end{equation}
where the summation on the right goes over the set of
$k$-tuples $\pi=(\pi_1,\dots,\pi_k)$ such that no proper subset of
$(1,\dots,n)$ is invariant under the group generated by the $\pi_j$. (Each
such $\pi$ corresponds to a connected graph consisting of edges beginning
at $p$ and ending at
$\pi_j(p)$, for all $p,j$.) Indeed,
\begin{equation}\label{t-h7}
 Q_n(z^1,\dots,z^n)=
\det A_{nkm}(z^1,\dots,z^n)
-{\sum_F}' 
\prod_{j=1}^l Q_{n_j}(z^{p_{j1}},\dots,z^{p_{jn_j}}), 
\end{equation}
where the summation on the right 
goes over all partitions $F=(F_1,\dots,F_l)$ with
$l\ge 2$. Using this equation,
we prove  (\ref{t-h6}) by induction (cf.\ the proof of Lemma
\ref{lem_t-h1}). The estimate (\ref {t-h4}) now follows from (\ref{t-h6})
and (\ref{ABC2}). \end {proof}  

\begin{lem}\label{lem_t-h3}
 The functions $ R_n(z^1,\dots,z^n)$ satisfy the estimate
\begin{equation}\label{t-h8}
 R_n(z^1,\dots,z^n)=O(d(z^1,\dots,z^n))\,,
\end{equation}
provided that $\min_{p\ne q}|z^p-z^q|\ge c>0$.
\end{lem}

\begin{proof} We have the identity
\begin{equation}\label{t-h9}
0=
{\sum_{G,H}}^{\rm conn}\,
\prod_{j=1}^l   Q_{n_j}(z^{p_{j1}},\dots,z^{p_{jn_j}})
\prod_{j=1}^{l'}  R_{m_j}(z^{p'_{j1}},\dots,z^{p'_{jm_j}})\,,\quad n\ge 2\,.
\end{equation}
The proof of this identity is the same as that of Lemma
\ref{lem_t-h1}. Indeed, the connected functions of $\det
A_{nkm} \,\displaystyle
\frac{1}{\det A_{nkm}}=1$ are equal to 0 (except that the 1-point connected
function equals 1); hence (\ref{t-h9}) follows. 

The identity (\ref{t-h9}) can
be rewritten as
\begin{equation}\label{t-h10}
\begin{aligned}
\det A_{nkm}(z^1,\dots&,z^n)\;R_n(z^1,\dots,z^n)\\
&=-{\sum_{G,H}}^{{\rm conn}'}\,
\prod_{j=1}^l   Q_{n_j}(z^{p_{j1}},\dots,z^{p_{jn_j}})
\prod_{j=1}^{l'}  R_{m_j}(z^{p'_{j1}},\dots,z^{p'_{jm_j}})\end{aligned}
\end{equation}
where the summation on the right goes over all mutually connected
pairs of partitions
$\{G,H\}$ with at least two elements in $H$ (i.e., $l'\ge 2$). Now the
estimate (\ref{t-h8}) follows by induction from Lemma \ref{lem_t-h2} and
identity (\ref{t-h10}). 
\end{proof}

Theorem \ref{cc} follows from Lemmas \ref{cchat2}, \ref{lem_t-h1}  and
\ref{lem_t-h3}.  The theorem yields the following more explicit estimate:

\begin{cor}\label{ccc} The connected correlation functions satisfy the
estimate
$$\wt T^\infty_{nkm}(z^1,\dots,z^n) = o\big(e^{-R^2/n}\big) \,,\quad R=
\max_{p,q}|z^p-z^q|\,,$$ provided that
$\min_{p\ne q}|z^p-z^q|\ge c>0$. \end{cor}

\begin{proof} We must show that
\begin{equation}\label{cccd} d(z^1,\dots,z^n)\le
o\big(e^{-R^2/n}\big)\,.\end{equation} Assume without loss of generality that
$|z^1-z^n|=R$.  Let $\gcal=(V,L)$ be an oriented connected graph with zero
boundary as in the definition of $d(z^1,\dots,z^n)$. Since $z^1$ and $z^n$ are
connected by a chain of loops in $\gcal$, we can choose disjoint sets of edges
$L',L''\subset L$ such that $L'$ forms a path starting at $z^1$ and ending at
$z^n$, and $L''$ forms a path starting at $z^n$ and ending at $z^1$.  This
means that there is a sequence $z^1=z^{i_1},z^{i_2},\dots,z^{i_{n'}}=z^n$ such
that $L'=\{l_1,\dots,l_{n'-1}\}$, where $l_j$ begins at
$z^{i_j}$ and ends at $z^{i_{j+1}}$. By removing any loops in $L'$, we can
assume that the $z^{i_j}$ are distinct and thus $n' \le n$.  A similar
description holds for $L''$.  Let $r_j=|z^{i_j}- z^{i_{j+1}}|$.  We note that
$$R\le \sum r_j \le \left((n'-1)\sum r_j^2\right)^{1/2},$$
where the second inequality is by Cauchy-Schwarz.  We then have
$$\prod_{l\in L'}
|z^{i(l)}-z^{f(l)}|^2e^{-|z^{i(l)}-z^{f(l)}|^2/2}=\prod_{j=1}^{n'-1}
r_j^2 e^{-r_j^2/2}
\le R^{2n'-2}e^{-\half \sum r_j^2} \le R^{2n'-2}e^{-R^2/(2n'-2)}\,,$$
and hence
\begin{equation}\label{path}\prod_{l\in L'}
|z^{i(l)}-z^{f(l)}|^2e^{-|z^{i(l)}-z^{f(l)}|^2/2}
\le R^{2n-2}e^{-R^2/(2n-2)}\,,\quad R\ge 1\,.
\end{equation}
The same inequality also holds for the product over the path $L''$.
Since each term of
the product in (\ref{cc4}) is less than 1, we then have
$$\prod_{l\in L}
|z^{i(l)}-z^{f(l)}|^2e^{-|z^{i(l)}-z^{f(l)}|^2/2}
\le \prod_{l\in L'\cup L''}
|z^{i(l)}-z^{f(l)}|^2e^{-|z^{i(l)}-z^{f(l)}|^2/2}
\le o\left(e^{-R^2/n}\right)\,.$$
Taking the supremum over all graphs, we obtain (\ref{cccd}).
\end{proof}

\begin{rem} The above proof gives the bound
\begin{equation} d(z^1,\dots,z^n)\le R^{4n-4} e^{-R^2/(n-1)}\,,\quad R\ge
1\,. \end{equation}
Hence we actually have the estimate
\begin{equation} \wt T^\infty_{nkm}(z^1,\dots,z^n) = O\big(R^{4n-4}
e^{-R^2/(n-1)}\big)\,,\quad \mbox{provided that }\ 
\min_{p\ne q}|z^p-z^q|\ge c>0\,. \label{ccc1}\end{equation}
Equation (\ref{ccc1}) implies Theorem \ref{shortrange} because of the inversion
formula (\ref{cc3}).
\end{rem}


\begin{thebibliography}{HHHH}

\bibitem[At]{At} M. F. Atiyah, {\it The Geometry and Physics of Knots}, 
Lezioni
Lincee, Cambridge Univ.\ Press, Cambridge, 1990.

\bibitem[Au1]{A} D. Auroux,  Asymptotically holomorphic families of
symplectic submanifolds, {\it Geom.\ Funct.\ Anal.} 7 (1997),
971--995.

\bibitem[Au2]{A.2} D. Auroux, Symplectic 4-manifolds as branched
coverings of $\CP^2$, {\it Invent.\ Math.} 139  (2000),  551--602.

\bibitem[AK]{A.3} D. Auroux and L. Katzarkov,
Branched coverings of $\CP^2$ and invariants of symplectic 4-manifolds
(preprint, 1999).


\bibitem[BFG]{BFG} M. Beals, C. Fefferman and R. Grossman, Strictly
pseudoconvex
domains in $\C^n$, {\it Bull. Amer.\ Math.\ Soc.} 8 (1983), 125--322.

\bibitem[BD]{BD} P. Bleher and X. Di, Correlations between zeros of a random
polynomial, {\it J. Stat.\ Phys.} 88 (1997), 269--305.


\bibitem[BSZ1]{BSZ1} P. Bleher, B. Shiffman and S. Zelditch,
Poincar\'e-Lelong approach to universality and scaling of correlations
between zeros, {\it Comm.\ Math.\ Phys.} 208 (2000), 771--785.

\bibitem[BSZ2]{BSZ2}  P. Bleher, B. Shiffman and S. Zelditch,  Universality
and scaling of correlations between zeros on complex manifolds, {\it
Invent.\
Math.}, to appear, http://xxx.lanl.gov/abs/math-ph/9904020.

\bibitem[BSZ3]{BSZ3}  P. Bleher, B. Shiffman and S. Zelditch,
Correlations between zeros in higher dimensions (preprint, 2000).

\bibitem[BBL]{BBL} E. Bogomolny, O. Bohigas, and P. Leboeuf, Quantum chaotic
dynamics and random polynomials, {\it J. Stat.\ Phys.} 85 (1996), 639--679.


\bibitem[BU1]{BU.1} D. Borthwich and A. Uribe, Almost complex
structures and
geometric
quantization, {\it Math.\ Res.\ Lett.} 3 (1996), 845--861.

\bibitem[BU2]{BU.2} D. Borthwick and A. Uribe, Nearly K\"ahlerian
embeddings of symplectic manifolds,
http://xxx.lanl.gov/abs/math.DG/9812041.

\bibitem[Bch]{Bouche} T. Bouche, Asymptotic results for Hermitian
line bundles over complex manifolds: the heat kernel approach.
Higher-dimensional complex varieties (Trento, 1994), 67--81, de
Gruyter, Berlin, 1996.

\bibitem[Bou]{Bou} L. Boutet de Monvel, Hypoelliptic operators with
double characteristics and
related pseudodifferential operators, {\it Comm.\ Pure and Appl.\ Math.}
27 (1974), 585- 639.

\bibitem[BG]{BG} L.  Boutet de Monvel and V.  Guillemin, {\it The Spectral
Theory of Toeplitz Operators}, Ann.\ Math.\ Studies 99, Princeton Univ.\
Press, Princeton, 1981.

\bibitem[BK]{BK} A. Boutet de Monvel and A. Khorunzhy,
On universality of the smoothed eigenvalue density of large random matrices,
{\it J. Phys. A\/} 32  (1999), L413--L417.

\bibitem[BS]{BS} L. Boutet de Monvel and J. Sj\"ostrand, Sur la
singularit\'e des noyaux de Bergman et de Szeg\"o, {\it Asterisque\/} 34--35
(1976), 123--164.

\bibitem[BZ]{BZ} E. Brezin and A. Zee, Universality of the correlations
between
eigenvalues of large random matrices, {\it Nucl.\ Phys.\ B\/} 402 (1993),
613-627.

\bibitem[Car]{CAR} J. Cardy, {\it Scaling and Renormalization in
Statistical Physics},
Cambridge Lecture Notes in Physics 5, Cambridge Univ.\ Press, Cambridge,
1996.

\bibitem[Cat]{Cat} D. Catlin, The Bergman kernel and a theorem of
Tian, in: {\it Analysis and Geometry in Several Complex
Variables\/}, G. Komatsu and M. Kuranishi, eds., Birkh\"auser,
Boston, 1999.

\bibitem[De]{D} P. A. Deift, {\it  Orthogonal polynomials and random
matrices:
a Riemann-Hilbert approach.} Courant Lecture Notes in Mathematics, Courant
Institute of Mathematical Sciences, New York, 1999.

\bibitem[Do1]{DON.1} S.  Donaldson, Symplectic submanifolds and
almost complex
geometry, {\it J. Diff.\  Geom.}  44 (1996), 666--705.

\bibitem[Do2]{DON.2} S. Donaldson, Lefschetz fibrations in symplectic
geometry, {\it Proceedings of the International Congress of
Mathematicians,\/} Vol.\ II (Berlin, 1998). 

\bibitem[EK]{EK} A. Edelman and E. Kostlan, How many zeros of a random
polynomial are real? {\it Bull.\ Amer.\ Math.\ Soc.} 32 (1995), 1--37.



\bibitem[Fe]{Fe} H. Federer, {\it Geometric Measure Theory\/}, Springer, New
York, 1969.

\bibitem[Fo]{Fo} G. B. Folland, {\it Harmonic Analysis in Phase Space},
Princeton University Press, Princeton (1989).

\bibitem[GJ]{GJ} J. Glimm and A. Jaffe, {\it Quantum Physics. A
Functional
Integral Point of View}, 2nd ed., Springer-Verlag, New York (1987).


\bibitem[GH]{GH} P.  Griffiths and J.  Harris, {\it Principles of Algebraic
Geometry}, Wiley-Interscience, New York, 1978.

\bibitem[GS]{GS} V. Guillemin and S. Sternberg, {\it Symplectic Techniques in
Physics},
Cambridge Univ. Press, Cambridge (1984).

\bibitem[GU]{GU} V. Guillemin and A. Uribe, Laplace operator on tensor
powers
of a line bundle, {\it Asympt.\ Anal.}  1 (1988), 105--113.

\bibitem[Hal]{Halp} B. I. Halperin, Statistical mechanics of
topological defects, in: {\it Physics of Defects, Les Houches
Session XXXV,\/} North-Holland
(1980).

\bibitem[Han]{Ha} J. H. Hannay, Chaotic analytic zero points: exact
statistics
for those of a random spin state, {\it J. Phys.\ A: Math.\ Gen.} 29 (1996),
101--105.

\bibitem[H\"o]{H} L. H\"ormander, {\it The Analysis of Linear
Partial Differential
Operators I}, Grund.\ Math.\ Wiss.\ 256, Springer-Verlag, N.Y.
(1983).

\bibitem[Kac]{Ka} M. Kac, On the average number of real roots of a random
algebraic equation, {\it Bull.\ Amer.\ Math.\ Soc.\ } 49 (1943), 314--320.

\bibitem[KS]{KS} N. Katz and P. Sarnak, {\it Random Matrices, Frobenius
Eigenvalues,
and Monodromy}, AMS Colloq.\ Pub.\ 45, Providence, RI (1998).

\bibitem[Kl]{Kl} S. Kleiman, Toward a numerical theory of ampleness,
{\it Ann.\ of Math.} 84 (1966), 293--344.

\bibitem[Le]{Le} P. Lelong, Int\'egration sur un ensemble analytique
complexe,
{\it Bull.\ Soc.\ Math.\ France\/} 85 (1957), 239--262.

\bibitem[MM]{MM} V. A. Malyshev and R. A. Minlos, {\it Gibbs Random
Fields. Cluster Expansions}, Kluwer Acad.\ Publ., Dordrecht, 1991. 

\bibitem[MelS]{MeS} A. Melin and J. Sj\"ostrand, Fourier integral
operators with complex valued phase functions,
Springer Lecture Notes 459, p. 120-233, Springer,  New York, 1975.

\bibitem[MenS]{MS} A. Menikoff and J. Sj\"ostrand, On the eigenvalues
of a class of hypoelliptic operators, {\it Math.\ Ann.} 235 (1978),
55--85.

\bibitem[Na]{Na} Y. Nakai, A criterion of an ample sheaf on a
projective scheme, {\it Amer.\ J. Math.} 85 (1963), 14--26.

\bibitem[Ne]{Ne} J. Neuheisel, The asymptotic distribution of nodal sets on
spheres, Ph.D.\ thesis, Johns Hopkins University, 2000.

\bibitem[NV]{NV} S. Nonnenmacher and A. Voros, Chaotic eigenfunctions in
phase space, {\it J. Stat.\ Phys.} 92 (1998), 431--518.

\bibitem[Ri]{Ri} S. O. Rice, Mathematical analysis of random noise, {\it
Bell System Tech.\ J.} 23 (1944), 282--332, and 24 (1945), 46--156;
reprinted in: Selected papers on noise and stochastic processes, Dover,
New York (1954), pp.\ 133--294.

\bibitem[RS]{RS} Z. Rudnick and P. Sarnak, The pair correlation function of
fractional parts of polynomials. {\it Comm.\ Math.\ Phys.} 194 (1998),
61--70


\bibitem[Sa]{Sa} P. Sarnak, Arithmetic quantum chaos, in {\it Israel
Mathematical
Conference Proc. Vol. 8} (1995), 183--236.

\bibitem[SSo]{SS} B.  Shiffman and A.  J.  Sommese, {\it Vanishing
Theorems on Complex Manifolds\/}, Progress in Math.\ 56, Birkh\"auser,
Boston, 1985. 


\bibitem[SZ1]{SZ} B. Shiffman and S. Zelditch, Distribution
of zeros of random and quantum chaotic sections of positive line bundles,
{\it Comm.\ Math.\ Phys.}   200 (1999),
661--683.

\bibitem[SZ2]{SZ2} B. Shiffman and S. Zelditch, Random almost holomorphic
sections of ample line bundles on symplectic manifolds,
http://xxx.lanl.gov/abs/math.SG/0001102.

\bibitem[SSm]{ShSm} M. Shub and S. Smale, Complexity of Bezout's theorem
II: Volumes and probabilities, in: {\it Computational Algebraic Geometry
(Nice, 1992)},  Progr.\ Math.\ 109, Birkh\"auser, Boston,
(1993), pp.\ 267--285.

\bibitem[Sik]{Sik} J. C. Sikorav,  Construction de sous-vari\'et\'es
symplectiques (d'apr\`es S. K. Donaldson et D. Auroux), S\'eminaire
Bourbaki. Vol.\ 1997/98. Ast\'erisque 252, (1998), 231--253.


\bibitem[Sin]{SI} Ya. G. Sinai, {\it Theory of Phase Transitions:
Rigorous Results},
Pregamon Press, New York, 1982.

\bibitem[Sj]{Sj} J. Sj\"ostrand, Density of resonances for strictly
convex analytic obstacles, {\it Can.\ J. Math.} 48 (1996),
397--447.

\bibitem[So]{So} A. Soshnikov,
Universality at the edge of the spectrum in Wigner random matrices,
{\it Comm.\ Math.\ Phys.} 207  (1999), 697--733.

\bibitem[St]{Stein} E. M. Stein, {\it Harmonic Analysis}, Princeton
University Press,
Princeton (1993).

\bibitem[Ti]{Ti} G.  Tian, On a set of polarized \kahler metrics on
algebraic
manifolds, {\it J.  Diff.\ Geometry\/} 32 (1990), 99--130.

\bibitem[TW]{TW} C. A. Tracy and H. Widom,
Universality of the distribution functions of random matrix theory. II,
http://xxx.lanl.gov/abs/math-ph/9909001.

\bibitem[Wo]{W} N. M. J. Woodhouse, {\it Geometric Quantization},
Clarendon Press,
Oxford, 1992.

\bibitem[Ze1]{Ze} S. Zelditch,  Szeg\"o kernels and a theorem of
Tian, {\it Int.\ Math.\ Res.\ Notices} 6 (1998), 317--331.

\bibitem[Ze2]{Ze.2} S. Zelditch, Level spacings for integrable
quantum maps in genus zero, {\it Comm.\ Math.\ Phys.} 196 (1998),
289--318.


\bibitem[ZZ]{ZZ} S. Zelditch and M. Zworski, Spacings between phase shifts
in a simple scattering problem, {\it Comm.\ Math.\ Phys.}
204 (1999), 709--729.



\bibitem[Zi]{Zinn} P. Zinn-Justin,
Universality of correlation functions of Hermitian random matrices in an
external field,
{\it Comm.\ Math.\ Phys.}  194 (1998),  631--650.

\end{thebibliography}
\end{document}